\documentstyle[10pt,epsf,epsfig,dp_delphititle]{dp_delphi}
\makeindex
\def\DpPaperGroup{EP}
\def\DpPaperRef{99--37}
\def\DpDate{8 March 1999}
\def\DpAuthors{DELPHI Collaboration}
\def\DpSubmit{(Submitted to E. Phys. J. C)}
\def\DpTitle{{Search for charginos nearly mass-degenerate with the lightest 
neutralino}}

\begin{document}
\makeatletter
\newcount\@tempcntc
\def\@citex[#1]#2{\if@filesw\immediate\write\@auxout{\string\citation{#2}}\fi
  \@tempcnta\z@\@tempcntb\m@ne\def\@citea{}\@cite{\@for\@citeb:=#2\do
    {\@ifundefined
       {b@\@citeb}{\@citeo\@tempcntb\m@ne\@citea\def\@citea{,}{\bf ?}\@warning
       {Citation `\@citeb' on page \thepage \space undefined}}%
    {\setbox\z@\hbox{\global\@tempcntc0\csname b@\@citeb\endcsname\relax}%
     \ifnum\@tempcntc=\z@ \@citeo\@tempcntb\m@ne
       \@citea\def\@citea{,}\hbox{\csname b@\@citeb\endcsname}%
     \else
      \advance\@tempcntb\@ne
      \ifnum\@tempcntb=\@tempcntc
      \else\advance\@tempcntb\m@ne\@citeo
      \@tempcnta\@tempcntc\@tempcntb\@tempcntc\fi\fi}}\@citeo}{#1}}
\def\@citeo{\ifnum\@tempcnta>\@tempcntb\else\@citea\def\@citea{,}%
  \ifnum\@tempcnta=\@tempcntb\the\@tempcnta\else
   {\advance\@tempcnta\@ne\ifnum\@tempcnta=\@tempcntb \else \def\@citea{--}\fi
    \advance\@tempcnta\m@ne\the\@tempcnta\@citea\the\@tempcntb}\fi\fi}
 
\makeatother
\begin{titlepage}
\pagenumbering{roman}
\CERNpreprint{\DpPaperGroup}{\DpPaperRef} 
\date{{\small\DpDate}} 
\title{\DpTitle} 
\address{\DpAuthors} 
\begin{shortabs} 
\noindent
%
\noindent

A search for charginos with masses close to the mass of the lightest neutralino
is reported, based on the data collected with the DELPHI detector at LEP from 
1995 to 1997 at centre-of-mass energies between 130 and 183 GeV.
The signature of a photon at high transverse momentum radiated from the initial
state reduces the two-photon background to acceptable rates, thus making the
mass differences between a few hundred MeV/c$^2$ and 3~GeV/c$^2$ detectable.
In very nearly degenerate scenarios, the lifetime of the chargino can be large
enough to produce either visible secondary vertices or decays outside the
detector; therefore, quasi-stable heavy charged particles and displaced
decay vertices were also searched for. No excess of events with respect to
the Standard Model expectations was observed, and limits in the plane of 
chargino-neutralino mass difference versus chargino mass are given.

\end{shortabs}
\vfill
\begin{center}
\DpSubmit \ 
\end{center}
\vfill
\clearpage
\headsep 10.0pt
\addtolength{\textheight}{10mm}
\addtolength{\footskip}{-5mm}
\begingroup
%
\newcommand{\DpName}[2]{\hbox{#1$^{\ref{#2}}$},\hfill}
\newcommand{\DpNameTwo}[3]{\hbox{#1$^{\ref{#2},\ref{#3}}$},\hfill}
\newcommand{\DpNameThree}[4]{\hbox{#1$^{\ref{#2},\ref{#3},\ref{#4}}$},\hfill}
\newskip\Bigfill \Bigfill = 0pt plus 1000fill
\newcommand{\DpNameLast}[2]{\hbox{#1$^{\ref{#2}}$}\hspace{\Bigfill}}
\small
\noindent
\DpName{P.Abreu}{LIP}
\DpName{W.Adam}{VIENNA}
\DpName{T.Adye}{RAL}
\DpName{P.Adzic}{DEMOKRITOS}
\DpName{Z.Albrecht}{KARLSRUHE}
\DpName{T.Alderweireld}{AIM}
\DpName{G.D.Alekseev}{JINR}
\DpName{R.Alemany}{VALENCIA}
\DpName{T.Allmendinger}{KARLSRUHE}
\DpName{P.P.Allport}{LIVERPOOL}
% ...
\normalsize
\endgroup
\titlefoot{Department of Physics and Astronomy, Iowa State
     University, Ames IA 50011-3160, USA
    \label{AMES}}
\titlefoot{Physics Department, Univ. Instelling Antwerpen,
     Universiteitsplein 1, BE-2610 Wilrijk, Belgium \\
     \indent~~and IIHE, ULB-VUB,
     Pleinlaan 2, BE-1050 Brussels, Belgium \\
     \indent~~and Facult\'e des Sciences,
     Univ. de l'Etat Mons, Av. Maistriau 19, BE-7000 Mons, Belgium
    \label{AIM}}
\titlefoot{Physics Laboratory, University of Athens, Solonos Str.
     104, GR-10680 Athens, Greece
    \label{ATHENS}}
\titlefoot{Department of Physics, University of Bergen,
     All\'egaten 55, NO-5007 Bergen, Norway
    \label{BERGEN}}
\titlefoot{Dipartimento di Fisica, Universit\`a di Bologna and INFN,
     Via Irnerio 46, IT-40126 Bologna, Italy
    \label{BOLOGNA}}
\titlefoot{Centro Brasileiro de Pesquisas F\'{\i}sicas, rua Xavier Sigaud 150,
     BR-22290 Rio de Janeiro, Brazil \\
     \indent~~and Depto. de F\'{\i}sica, Pont. Univ. Cat\'olica,
     C.P. 38071 BR-22453 Rio de Janeiro, Brazil \\
     \indent~~and Inst. de F\'{\i}sica, Univ. Estadual do Rio de Janeiro,
     rua S\~{a}o Francisco Xavier 524, Rio de Janeiro, Brazil
    \label{BRASIL}}
\titlefoot{Comenius University, Faculty of Mathematics and Physics,
     Mlynska Dolina, SK-84215 Bratislava, Slovakia
    \label{BRATISLAVA}}
\titlefoot{Coll\`ege de France, Lab. de Physique Corpusculaire, IN2P3-CNRS,
     FR-75231 Paris Cedex 05, France
    \label{CDF}}
\titlefoot{CERN, CH-1211 Geneva 23, Switzerland
    \label{CERN}}
\titlefoot{Institut de Recherches Subatomiques, IN2P3 - CNRS/ULP - BP20,
     FR-67037 Strasbourg Cedex, France
    \label{CRN}}
\titlefoot{Institute of Nuclear Physics, N.C.S.R. Demokritos,
     P.O. Box 60228, GR-15310 Athens, Greece
    \label{DEMOKRITOS}}
\titlefoot{FZU, Inst. of Phys. of the C.A.S. High Energy Physics Division,
     Na Slovance 2, CZ-180 40, Praha 8, Czech Republic
    \label{FZU}}
\titlefoot{Dipartimento di Fisica, Universit\`a di Genova and INFN,
     Via Dodecaneso 33, IT-16146 Genova, Italy
    \label{GENOVA}}
\titlefoot{Institut des Sciences Nucl\'eaires, IN2P3-CNRS, Universit\'e
     de Grenoble 1, FR-38026 Grenoble Cedex, France
    \label{GRENOBLE}}
\titlefoot{Helsinki Institute of Physics, HIP,
     P.O. Box 9, FI-00014 Helsinki, Finland
    \label{HELSINKI}}
\titlefoot{Joint Institute for Nuclear Research, Dubna, Head Post
     Office, P.O. Box 79, RU-101 000 Moscow, Russian Federation
    \label{JINR}}
\titlefoot{Institut f\"ur Experimentelle Kernphysik,
     Universit\"at Karlsruhe, Postfach 6980, DE-76128 Karlsruhe,
     Germany
    \label{KARLSRUHE}}
\titlefoot{Institute of Nuclear Physics and University of Mining and Metalurgy,
     Ul. Kawiory 26a, PL-30055 Krakow, Poland
    \label{KRAKOW}}
\titlefoot{Universit\'e de Paris-Sud, Lab. de l'Acc\'el\'erateur
     Lin\'eaire, IN2P3-CNRS, B\^{a}t. 200, FR-91405 Orsay Cedex, France
    \label{LAL}}
\titlefoot{School of Physics and Chemistry, University of Lancaster,
     Lancaster LA1 4YB, UK
    \label{LANCASTER}}
\titlefoot{LIP, IST, FCUL - Av. Elias Garcia, 14-$1^{o}$,
     PT-1000 Lisboa Codex, Portugal
    \label{LIP}}
\titlefoot{Department of Physics, University of Liverpool, P.O.
     Box 147, Liverpool L69 3BX, UK
    \label{LIVERPOOL}}
\titlefoot{LPNHE, IN2P3-CNRS, Univ.~Paris VI et VII, Tour 33 (RdC),
     4 place Jussieu, FR-75252 Paris Cedex 05, France
    \label{LPNHE}}
\titlefoot{Department of Physics, University of Lund,
     S\"olvegatan 14, SE-223 63 Lund, Sweden
    \label{LUND}}
\titlefoot{Universit\'e Claude Bernard de Lyon, IPNL, IN2P3-CNRS,
     FR-69622 Villeurbanne Cedex, France
    \label{LYON}}
\titlefoot{Univ. d'Aix - Marseille II - CPP, IN2P3-CNRS,
     FR-13288 Marseille Cedex 09, France
    \label{MARSEILLE}}
\titlefoot{Dipartimento di Fisica, Universit\`a di Milano and INFN,
     Via Celoria 16, IT-20133 Milan, Italy
    \label{MILANO}}
\titlefoot{Niels Bohr Institute, Blegdamsvej 17,
     DK-2100 Copenhagen {\O}, Denmark
    \label{NBI}}
\titlefoot{NC, Nuclear Centre of MFF, Charles University, Areal MFF,
     V Holesovickach 2, CZ-180 00, Praha 8, Czech Republic
    \label{NC}}
\titlefoot{NIKHEF, Postbus 41882, NL-1009 DB
     Amsterdam, The Netherlands
    \label{NIKHEF}}
\titlefoot{National Technical University, Physics Department,
     Zografou Campus, GR-15773 Athens, Greece
    \label{NTU-ATHENS}}
\titlefoot{Physics Department, University of Oslo, Blindern,
     NO-1000 Oslo 3, Norway
    \label{OSLO}}
\titlefoot{Dpto. Fisica, Univ. Oviedo, Avda. Calvo Sotelo
     s/n, ES-33007 Oviedo, Spain
    \label{OVIEDO}}
\titlefoot{Department of Physics, University of Oxford,
     Keble Road, Oxford OX1 3RH, UK
    \label{OXFORD}}
\titlefoot{Dipartimento di Fisica, Universit\`a di Padova and
     INFN, Via Marzolo 8, IT-35131 Padua, Italy
    \label{PADOVA}}
\titlefoot{Rutherford Appleton Laboratory, Chilton, Didcot
     OX11 OQX, UK
    \label{RAL}}
\titlefoot{Dipartimento di Fisica, Universit\`a di Roma II and
     INFN, Tor Vergata, IT-00173 Rome, Italy
    \label{ROMA2}}
\titlefoot{Dipartimento di Fisica, Universit\`a di Roma III and
     INFN, Via della Vasca Navale 84, IT-00146 Rome, Italy
    \label{ROMA3}}
\titlefoot{DAPNIA/Service de Physique des Particules,
     CEA-Saclay, FR-91191 Gif-sur-Yvette Cedex, France
    \label{SACLAY}}
\titlefoot{Instituto de Fisica de Cantabria (CSIC-UC), Avda.
     los Castros s/n, ES-39006 Santander, Spain
    \label{SANTANDER}}
\titlefoot{Dipartimento di Fisica, Universit\`a degli Studi di Roma
     La Sapienza, Piazzale Aldo Moro 2, IT-00185 Rome, Italy
    \label{SAPIENZA}}
\titlefoot{Inst. for High Energy Physics, Serpukov
     P.O. Box 35, Protvino, (Moscow Region), Russian Federation
    \label{SERPUKHOV}}
\titlefoot{J. Stefan Institute, Jamova 39, SI-1000 Ljubljana, Slovenia
     and Laboratory for Astroparticle Physics,\\
     \indent~~Nova Gorica Polytechnic, Kostanjeviska 16a, SI-5000 Nova Gorica, Slovenia, \\
     \indent~~and Department of Physics, University of Ljubljana,
     SI-1000 Ljubljana, Slovenia
    \label{SLOVENIJA}}
\titlefoot{Fysikum, Stockholm University,
     Box 6730, SE-113 85 Stockholm, Sweden
    \label{STOCKHOLM}}
\titlefoot{Dipartimento di Fisica Sperimentale, Universit\`a di
     Torino and INFN, Via P. Giuria 1, IT-10125 Turin, Italy
    \label{TORINO}}
\titlefoot{Dipartimento di Fisica, Universit\`a di Trieste and
     INFN, Via A. Valerio 2, IT-34127 Trieste, Italy \\
     \indent~~and Istituto di Fisica, Universit\`a di Udine,
     IT-33100 Udine, Italy
    \label{TU}}
\titlefoot{Univ. Federal do Rio de Janeiro, C.P. 68528
     Cidade Univ., Ilha do Fund\~ao
     BR-21945-970 Rio de Janeiro, Brazil
    \label{UFRJ}}
\titlefoot{Department of Radiation Sciences, University of
     Uppsala, P.O. Box 535, SE-751 21 Uppsala, Sweden
    \label{UPPSALA}}
\titlefoot{IFIC, Valencia-CSIC, and D.F.A.M.N., U. de Valencia,
     Avda. Dr. Moliner 50, ES-46100 Burjassot (Valencia), Spain
    \label{VALENCIA}}
\titlefoot{Institut f\"ur Hochenergiephysik, \"Osterr. Akad.
     d. Wissensch., Nikolsdorfergasse 18, AT-1050 Vienna, Austria
    \label{VIENNA}}
\titlefoot{Inst. Nuclear Studies and University of Warsaw, Ul.
     Hoza 69, PL-00681 Warsaw, Poland
    \label{WARSZAWA}}
\titlefoot{Fachbereich Physik, University of Wuppertal, Postfach
     100 127, DE-42097 Wuppertal, Germany
    \label{WUPPERTAL}}
\titlefoot{On leave of absence from IHEP Serpukhov
    \label{MILAN-SERPOU}}
\titlefoot{Now at University of Florida
    \label{FLORIDA}}
\addtolength{\textheight}{-10mm}
\addtolength{\footskip}{5mm}
\clearpage
\headsep 30.0pt
\end{titlepage}
%
\pagenumbering{arabic} 
\setcounter{footnote}{0} %
\large
%


\def\cguz{\tilde\chi_1^0}
\def\cgdz{\tilde\chi_2^0}
\def\cgpm{\tilde\chi_1^{\pm}}
\def\cgdpm{\tilde\chi_2^{\pm}}
\def\cgup{\tilde\chi_1^+}
\def\cgum{\tilde\chi_1^-}
\def\snu{\tilde\nu}
\def\mchi{M_{\tilde \chi}}
\def\lchi{\lambda_{\tilde \chi}}
\def\tchi{\tau_{\tilde \chi}}
\def\mcgpm{M_{\cgpm}}
\def\mcgup{M_{\cgup}}
\def\mcguz{M_{\cguz}}
\def\mcgdz{M_{\cgdz}}
\def\msnu{M_{\snu}}
\def\epem{{\mathrm{e}}^+{\mathrm{e}}^-}
\def\Zzero{\mathrm{Z}}

\def\deg{^\circ}
\def\cms{centre-of-mass}
\def\mmin{\mathrm{min}}
\def\eff{\varepsilon}
\def\esel{\eff^{\mathrm{sel}}}
\def\etrg{\eff^{\mathrm{trg}}}
\newcommand{\lsim}{\;\raisebox{-0.9ex}{$\textstyle\stackrel{\textstyle<}{\sim}$}\;}

\def\ecms{E_{\mathrm{cms}}}
\def\evis{E_{\mathrm{vis}}}
\def\mopp{M_{\mathrm{opp}}}

\def\gamgam{\gamma\gamma}

\def\DMP{\Delta M ^{\pm}}
\def\DMZ{\Delta M ^0}

\def\nsig{N_{\mathrm{sig}}}
\def\nexp{N_{\mathrm{exp}}}

\def\gevc2{GeV/$c^2$}
\def\mevc2{MeV/$c^2$}
\def\tevc2{TeV/$c^2$}

\section{Introduction}

Supersymmetry (SUSY) \cite{susy} is an appealing theory which answers some
well-known outstanding questions of the Standard Model (SM), at the expense of
introducing supersymmetric partners of the known particles (sparticles).
The most well-studied example of supersymmetry is the minimal supersymmetric
extension of the SM (the MSSM). 
If the $R$-parity quantum number is conserved, as often assumed,
there must exist a lightest supersymmetric particle
(LSP) which is stable and remains after any SUSY decay chain. 
Such an LSP is expected to be neutral and weakly interacting \cite{nuclei}.
                                            The usual way of searching for SUSY
particles heavier than the LSP in $\epem$ interactions is therefore to look for
visible particles accompanied by the missing energy carried away by two or more
LSP's. This works whenever the mass difference between the produced sparticle
and the LSP is large enough to leave some sizeable amount of energy for the
visible final state particles. Typically, the searches carried out so far at
LEP in the different SUSY channels go down to mass differences of a few \gevc2.
Previous DELPHI searches at LEP2 have resulted in limits on the production of
charginos, neutralinos, sleptons and 
$\tilde {\mathrm b}$ and $\tilde {\mathrm t}$ squarks in the
MSSM, valid when the mass difference between the sparticles searched for and
the LSP (usually the lightest neutralino)
is above $3$ to $5$ \gevc2 \cite{delphisusy,delphicharg}. For smaller
mass differences, the only limits available so far are those derived from the
precise measurement of the $\Zzero$ width at LEP1; in particular charginos
lighter than about $M_{\Zzero}/2$ are excluded, irrespective of their field
content \cite{pdgnew,lep1}.

The search for charginos and neutralinos is essential to constrain SUSY. It is
therefore of paramount importance to make sure that they have not been missed
because of possible small mass differences amongst them. Such small mass
differences are rather unlikely in the MSSM if the masses of the gauginos are
assumed to be all the same at some grand unified scale, as expected in
supergravity (SUGRA) models. In such models the lightest chargino and the two
lightest neutralinos can have nearly the same mass only if those gaugino masses
are unnaturally large (above 1~\tevc2). However, since no direct experimental
support for those models has been found so far, it is reasonable to also
consider models without the SUGRA assumptions. In particular, there are
interesting theoretical string-motivated scenarios which explicitly prefer the 
non-unification of the gaugino masses at the GUT scale \cite{string} and in
which is quite likely that the lightest chargino and the lightest neutralino
have nearly equal masses \cite{cdg1}.

The region of low mass difference is experimentally challenging.
If the chargino-neutralino mass difference $\DMP = \mcgpm - \mcguz$ is below
the mass of the pion, the lifetime of the chargino can be so long that it passes
through the entire detector before decaying. In DELPHI this can be covered
by a search for heavy charged particles identified with the Cherenkov
detectors and/or because of their anomalously high ionization in the gas 
chambers. Mass differences of a few hundred \mevc2 may be observed by looking
for reconstructed secondary vertices from a chargino decay inside the detector
but significantly displaced from the main interaction point.

When $\DMP$ increases, the chargino decay length becomes so short that the
decay vertex can hardly be distinguished from the production one. However, as
long as $\DMP$ remains below a few \gevc2, the visible particles carry only a
small fraction of the parent energy.
A minimum visible transverse momentum is usually required to reject two-photon
interactions. This may result in an almost complete loss of efficiency for
chargino decays. On the other hand, some transverse momentum requirement is
necessary because the two-photon cross-section is orders of magnitude higher
than any signal searched for at LEP2. Here a suggestion by Chen, Drees and
Gunion \cite{cdg2} is applied to search for these charginos at low $\DMP$.
If one considers the events accompanied by a hard photon from Initial State
Radiation (ISR), then the two-photon background can be kept small by choosing
the photon transverse energy to be greater than
\begin{equation}
\label{eq:etgmin}
  (E_{\gamma}^T)^{\mmin} =
         {\sqrt s} \cdot {\sin \theta_{\mmin}  \over
                                     1 + \sin \theta_{\mmin}    }
      \,\,\,\,\,\,\, ,
\end{equation}
where $\theta_{\mmin}$ is the lowest polar angle accessible in the detector.
If an ISR photon with a transverse energy above $(E_{\gamma}^T)^{\mmin}$
is radiated from a two-photon event then, typically, one of the final state
electrons, which usually escape undetected in the beam pipe, should be
deflected at an angle larger than $\theta_{\mmin}$, thus allowing the
identification of the event as background. Such a selection gives a low
efficiency since only a small fraction of the SUSY events have an ISR photon
with $E_{\gamma}^T > (E_{\gamma}^T)^{\mmin}$.
On the other hand, the presence of the high energy photon in the detector
substantially increases the otherwise low trigger efficiency for these decays
with only a few soft visible particles.

This paper first explores the feasibility of a search at LEP2 for charginos or
second lightest neutralinos nearly mass-degenerate with the lightest neutralino.
In the case of the second lightest neutralino (for which the relevant mass
difference is $\DMZ = \mcgdz - \mcguz$) it will be shown that a search at LEP2
is either impossible or quite difficult, at least for most of the values of 
$\DMZ$ of interest here.
Instead, a search for mass-degenerate charginos was found to be feasible, and
was realized using the data collected by the DELPHI experiment. For sensitivity
in the case of long chargino lifetimes two alternative searches were used, as
described in section \ref{par:longliv}: one is the search for heavy stable
charged particles described in \cite{himass}, and the other is a modified
version of a search for secondary vertices \cite{kinks} with reconstructed
incoming and outgoing tracks (kinks). The search which exploits the ISR 
signature to cover the mass differences between $0.3$ and $3$~\gevc2 was
specifically designed for this work and is discussed in 
section~\ref{par:ISRTAG}.

\section{Data samples and event generators}
\label{par:data}

A detailed description of the working DELPHI detector can be found in 
\cite{delphidet}. The trajectories of the charged particles are reconstructed
in the $1.2$~T magnetic field by a system of cylindrical tracking chambers.
The most relevant for the analyses reported here are the Silicon Tracker, the 
Inner Detector (ID), and the Time Projection Chamber (TPC).
The Silicon Tracker is composed of the Vertex Detector (VD) in the barrel and
the ministrips and pixels of the Very Forward Tracker (VFT) at low polar
angles ($\theta$); it covers the range between $10\deg$ and $170\deg$ in
$\theta$ and radii down to $6.3$~cm from the beam. The ID covers polar angles
down to $15\deg$ ($165\deg$). The TPC tracks particles between the radii of 
29 and 122~cm, with at least three pad rows crossed if $\theta$ is between 
$20\deg$ and $160\deg$.
The electromagnetic calorimeters are the High density Projection Chamber
(HPC) in the barrel ($40\deg<\theta<140\deg$), the Forward ElectroMagnetic
Calorimeter (FEMC) in the forward regions ($11\deg<\theta<36\deg$ and
$144\deg<\theta<169\deg$) and the Small angle TIle Calorimeter (STIC) in the
very forward part (down to $1.66\deg$ from the beam axis). In front of the
STIC, which is also the luminometer of DELPHI, have been placed two planes
of scintillators (Veto Counters \cite{sticveto}) used to 
detect charged particles which enter the calorimeter.
Excellent particle identification is provided by the Ring Imaging CHerenkov
(RICH) detectors, equipped with two different radiators (liquid and gas)
with different refractive index, and thus different momentum thresholds.

During the high energy runs of LEP in 1995-97, DELPHI collected data at
\cms\ energies of approximately $130$, $136$, $161$, $172$ and $183$ GeV.
Only the runs in which the relevant subdetectors worked correctly were taken
into account for each analysis.
The luminosities used at the different energies in the analyses considered
here are approximately 6, 6, 10, 10 and 54 pb$^{-1}$ respectively. The data at
$130$ and $136$ GeV where not searched for kinks; in the search for soft
particles accompanied by ISR only 50~pb$^{-1}$ of the run at 183 GeV could be
used, mainly because of some temporary degradation of the quality of the data
collected by the HPC and FEMC calorimeters, which are fundamental for that
analysis.

To evaluate the signal efficiency and the background contamination, events were
generated using several different programs, all relying on JETSET 7.4
\cite{jetset} for quark fragmentation. All the events generated were passed
through a complete simulation of the DELPHI detector \cite{delsim}, and then
processed in exactly the same way as the real data.

The program SUSYGEN \cite{susygen}, which includes initial and final state
photon radiation, was used to simulate all signal events.
The implementation of the decay widths (i.e. branching ratios and lifetimes)
in SUSYGEN at low $\DMP$ has been modified in order to reproduce the results
of the analytical calculations reported in \cite{cdg1} and \cite{thomas}.

For the Standard Model background, several samples of the different final
states were generated with statistics which were typically well above those
expected (although some of the two-photon samples, especially at the lowest
\cms\ energies studied here, were originally simulated with statistics only
slightly higher than the one expected).
Annihiliation of $\epem$ into a virtual Z/$\gamma$, including ISR, were 
generated with PYTHIA \cite{jetset}. This generator was also used for 
four-fermion processes.
For the two-photon collisions, the generator of Berends, Daverveldt and
Kleiss (BDK) \cite{bdk} was used for the leptonic final states.
In this generator the $(\epem)\mu^+\mu^-$ and $(\epem)\tau^+\tau^-$ final
states include the simulation of the transverse momentum of the ISR photon, 
while in $(\epem)\epem$ only collinear ISR is implemented. In the two-photon 
interactions leading to hadronic final states, the QCD and VDM parts were 
simulated with the TWOGAM \cite{twogam} generator in which ISR is generated
without transverse momentum. BDK, with visible ISR, was used for the QPM part.
Although not used for the computation of the background, QPM events were also
generated using the TWOGAM generator, to study the differences between the two
programs, in particular effects related to the transverse momentum of the ISR
photons. The absence of these effects in some of the available two-photon
samples, implies that the simulation underestimates the background from ISR
in two-photon events.

\section{Chargino and neutralino production and decay at low $\Delta M$}
\label{par:scen}

The higgsino and gaugino sector of the MSSM can be described in terms
of four parameters: the ratio $\tan \beta$ of the two Higgs vacuum expectation
values, the Higgs mixing parameter $\mu$, the $SU(2)$ gaugino mass $M_2$ and
the $U(1)$ gaugino mass $M_1$. In the models with gaugino mass unification
at the GUT scale, there is a relation between $M_1$ and $M_2$
\begin{equation}
\label{eq:unif}
     M_1 = { 5 \over 3 } \tan ^2 \theta_W \cdot M_2
           \simeq 0.5 \cdot M_2   \,\,\,\,\,\,  .
\end{equation}
However, it has been already anticipated that this unification is not strictly
necessary in the theory and there are several models without it, in particular
the SUSY-string scenario proposed in \cite{cdg1}. The definition
\begin{equation}
\label{eq:rf}
     M_1 = R_f \cdot { 5 \over 3 } \tan ^2 \theta_W \cdot M_2 \,\,\,\,\,\, ,
\end{equation}
will be used, so that any value of $R_f$ different from $1$ indicates how
much the model deviates from the gaugino mass unification hypothesis.

In the MSSM there are two charginos ($\cgpm$ and $\cgdpm$). These mass
eigenstates are linear combinations of the two interaction eigenstates, the
wino and the charged higgsino. There are also four neutralinos, linear
combinations of the neutral interaction eigenstates. In the following it
will be assumed that the lightest neutralino is the LSP.

The lightest chargino gets almost the same mass as such an LSP in two cases
\cite{cdg1}:
\begin{enumerate}
\item{Low $|\mu|$, large $M_{1,2}$ scenario: $\cguz $ and $\cgup $ are
      both higgsino-like and nearly degenerate, with masses $\sim |\mu|$;}
\item{High $|\mu|$, low $M_2$ scenario: the $\cguz $ and the $\cgup $ are both
      gaugino-like and nearly degenerate with masses $\sim M_2$. In this
      scenario, in order to have $\DMP$ around $1$ \gevc2 or smaller,
      $R_f$ in eq.~(\ref{eq:rf}) must be larger than or equal to $2$.}
\end{enumerate}
In the first scenario, the second lightest neutralino is also almost
mass-degenerate with the lightest neutralino, with a mass splitting which is
sligtly larger than that of the lightest chargino. The same is true also for
the second scenario, but only for $R_f \simeq 2$.

\subsection{Cross-sections}

\begin{figure}[tbh]
\centerline{
\epsfxsize=10cm\epsffile{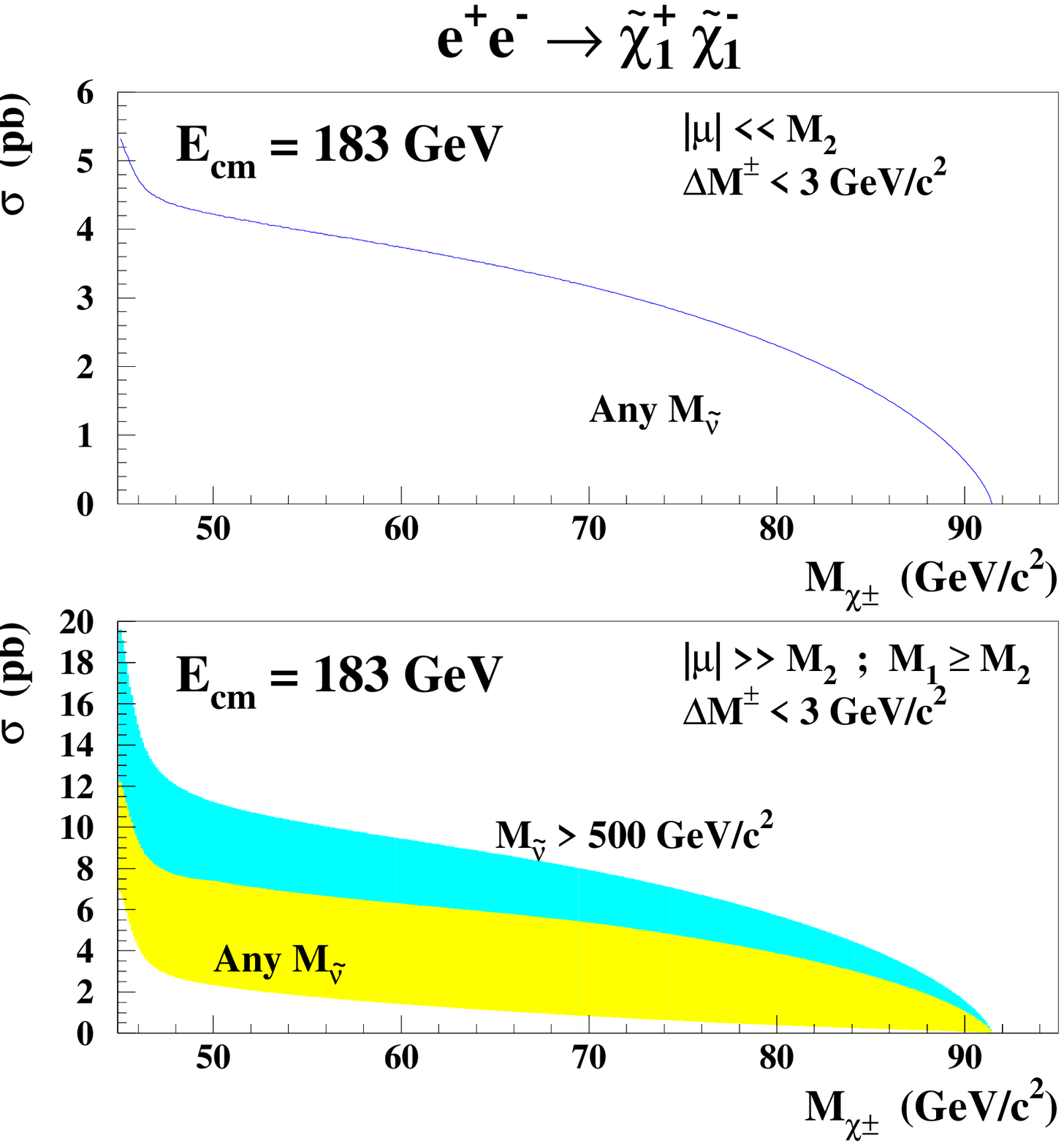} }
\caption[]{ Predicted $\epem \to \cgup \cgum$ cross-sections at the
   \cms\ energy of 183 GeV as a function of the mass of the chargino.
   The upper plot refers to the $|\mu| \ll M_2$ higgsino-like scenario;
   the lower plot to the $M_2 \ll |\mu|$ gaugino-like scenario.
   The widths of the bands allow for a variation
   of $M_1$, $M_2$ and $\mu$ so that $0<\DMP<3$~\gevc2; $1<\tan{\beta}<50$;
   $M_2 \le 2 M_1 \le 10M_2$;
   $\msnu>\mcgpm$ (the upper part of the gaugino band displays
   separately the points corresponding to $\msnu>500$~\gevc2).
                     }
\label{fig:xps}
\end{figure}

\begin{figure}[tbh]
\centerline{
\epsfxsize=10cm\epsffile{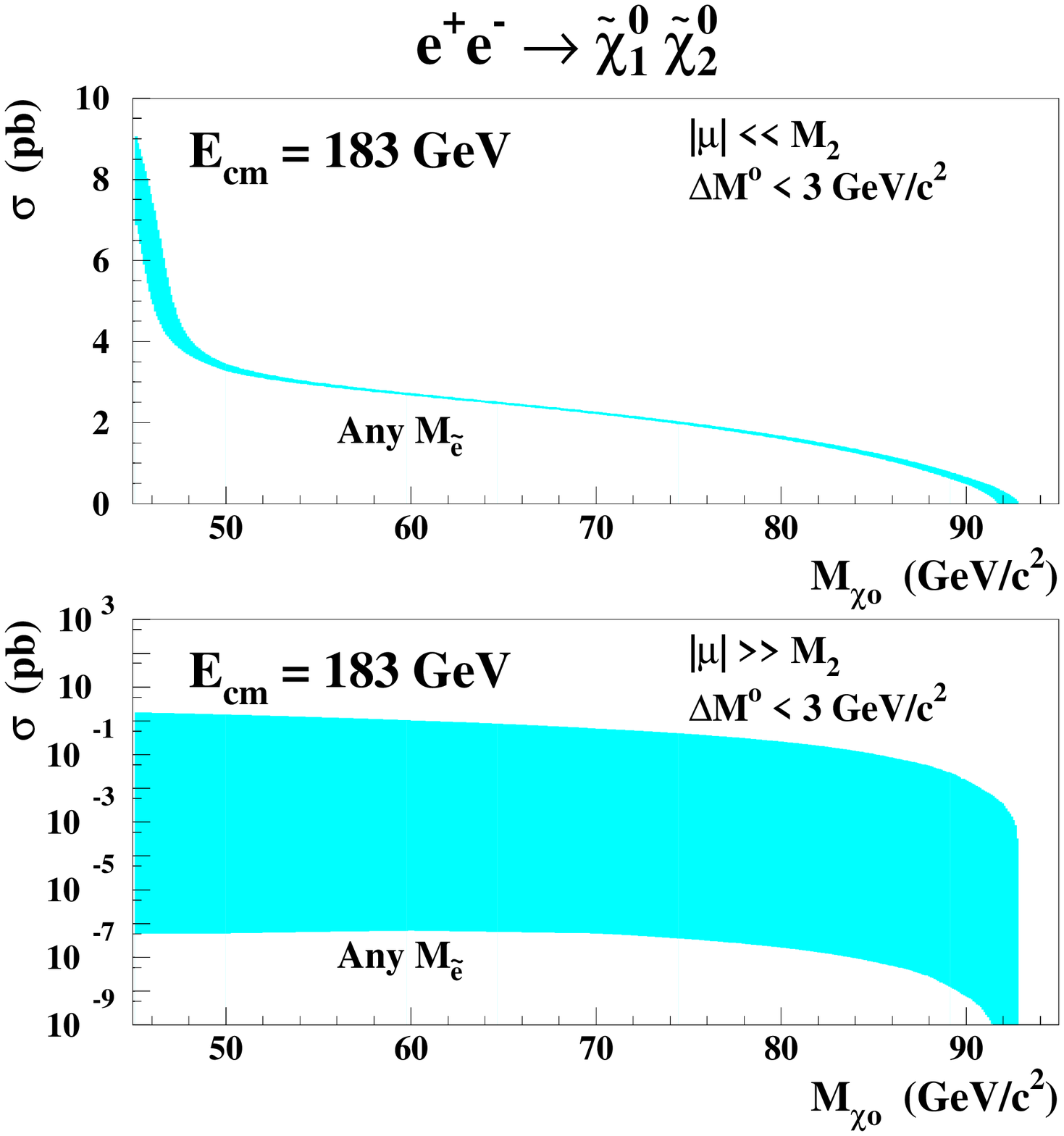} }
\caption[]{ Predicted $\epem \to \cguz \cgdz$ cross-sections at the \cms\
   energy of 183 GeV and as a function of the $\cgdz$ mass. The upper plot
   refers to $|\mu| \ll M_2$  higgsino-like scenario; the lower plot refers to
   the $M_2 \ll |\mu|$ gaugino like scenario. The width  of the bands allows
   for a variation of $M_2$, $M_1$ and $\mu$ so that for any given mass of the
   second neutralino $0<\DMZ<3$ \gevc2; $1<\tan{\beta}<50$;
   $1<R_f<10$ ($R_f\simeq 2$ for the gaugino); $M_{\tilde e}>\mcgdz$ .
                     }
\label{fig:xzs}
\end{figure}

The predicted $\epem\to\cgup \cgum $ and $\epem\to\cguz \cgdz $ cross-sections
\cite{susygen} at the \cms\ energy of 183 GeV are shown in figures
\ref{fig:xps} and \ref{fig:xzs}, respectively, as functions of the $\cgpm$ and
$\cgdz$ masses. That value of the \cms\ energy is taken as example, since the
behaviour is similar for all energies studied. In both figures, the upper plot
refers to the higgsino cross-section and the lower plot to the gaugino one.

For the neutralinos, the $\epem\to\cguz \cguz $  and $\epem\to\cgdz\cgdz$
cross-sections are much smaller than the $\epem\to\cguz\cgdz$ one, shown in
figure~\ref{fig:xzs}. This last process is the only one considered in the
following for the production of neutralinos at LEP2.

The widths of the bands arise from a variation of $\tan \beta$ from $1$ to
$50$, $M_2$ and $|\mu|$ below 100~\tevc2, $R_f$ between 1 and 10. The higgsino
cross-sections are quite stable when varying $R_f$, and also the values assumed
by the charged gaugino cross-sections do not deviate by more than $5\%$ if
$R_f$ moves in that range.
In the chargino production, the exchange of sfermions in the {\sl t}-channel
interferes destructively with the {\sl s}-channel production, lowering the
gaugino cross-section at small values of $\msnu$. On the contrary, in the
neutralino production the interference is constructive and the gaugino
cross-sections are enhanced at low $M_{\tilde e}$. In all the figures the mass
of the relevant scalar lepton has been varied from $\mchi$ to $1$ \tevc2.

The chargino cross-section, in the approximation of large scalar masses,
is roughly three times larger in the gau\-gi\-no-like sce\-na\-rio than
in the higgs\-ino-li\-ke one. On the contrary, the $\epem\to\cguz\cgdz$
cross-section in the gaugino-like scenario is typically several orders of
magnitude smaller than in the higgsino one, certainly out of reach of the
luminosity planned at LEP2.

\subsection{Branching ratios and lifetimes}
\label{par:theo}

The partial decay widths of the chargino in SUSYGEN have been modified
in order to account for the decays into a neutralino and one, two or three
pions, according to analytical calculations \cite{cdg1,thomas}.
These calculations use the form factors of the low mass hadronic resonances to
determine the hadronic width for $0<\DMP<2$~\gevc2. This treatment has been
verified \cite{cdg1} to describe correctly the $\tau$ decays, i.e. decays
with $\Delta M =m_{\tau}$.

\begin{figure}[htbp]
\centerline{
\epsfxsize=12cm\epsffile{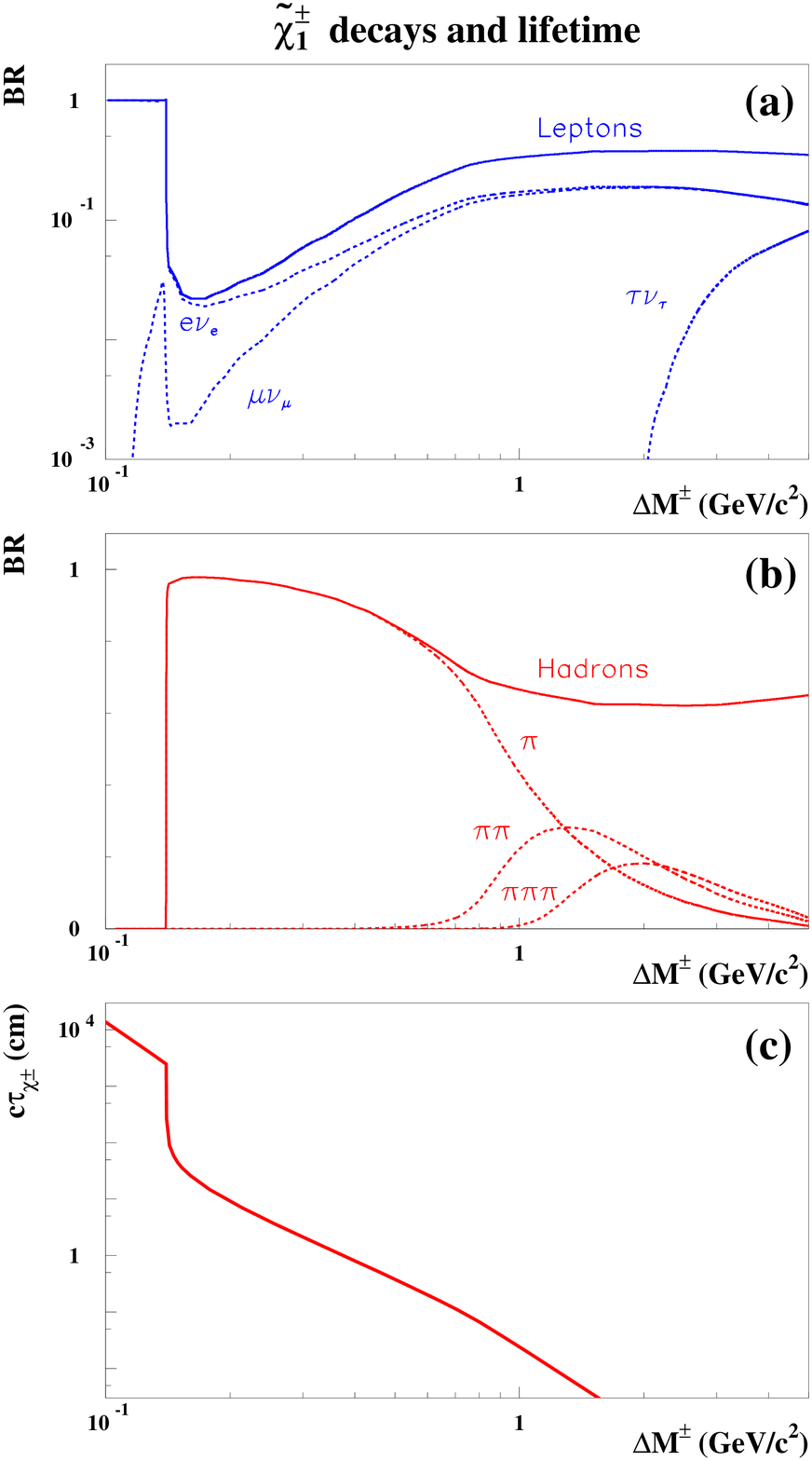} }
\caption[]{ Predicted (a) leptonic and (b) hadronic branching ratios and (c)
   lifetime of a $50$ \gevc2 chargino (higgsino) decaying into a $\cguz$
   plus standard particles, as a function of the mass difference between the
   $\cgup$ and the $\cguz$. Similar decay modes are predicted for the gaugino
   scenario in the approximation of large $\msnu$, which makes the contribution
   of the $\snu$ mediated decays negligible. For smaller $\msnu$, the leptonic
   decays of the charged gaugino are enhanced and the lifetime gets shorter.
                     }
\label{fig:brch}
\end{figure}

Figure \ref{fig:brch}a and b show the leptonic and hadronic branching ratios
computed for a 50~\gevc2 charged higgsino with a mass between 100~\mevc2 and
5~\gevc2 above that of the lightest neutralino. In the plot of the hadronic
modes, the decays into a $\cguz$ and one, two or three $\pi$, which contribute
to the total  $\cgup\to\cguz {\mathrm{q}} \bar {\mathrm{q}}$ width, are shown
separately.
For a gaugino, the leptonic width is enhanced for low $\msnu$ because of the
sneutrino mediated decays, and this was taken into account in the analysis.

Figure \ref{fig:brch}c shows the lifetime of the same chargino as a function of
$\DMP$. The figure clearly shows the step caused by the onset of the dominating
$\cgup \to \cguz \pi^+$ two body decay. The chargino lifetime and branching
ratios depend strongly on $\DMP$ and relatively little on the SUSY scenario,
as long as scalar exchange can be neglected. For low $\msnu$, charged gauginos
get a shorter lifetime, and this was also taken into account in the analysis.

\begin{figure}[htb]
\centerline{
\epsfxsize=12cm\epsffile{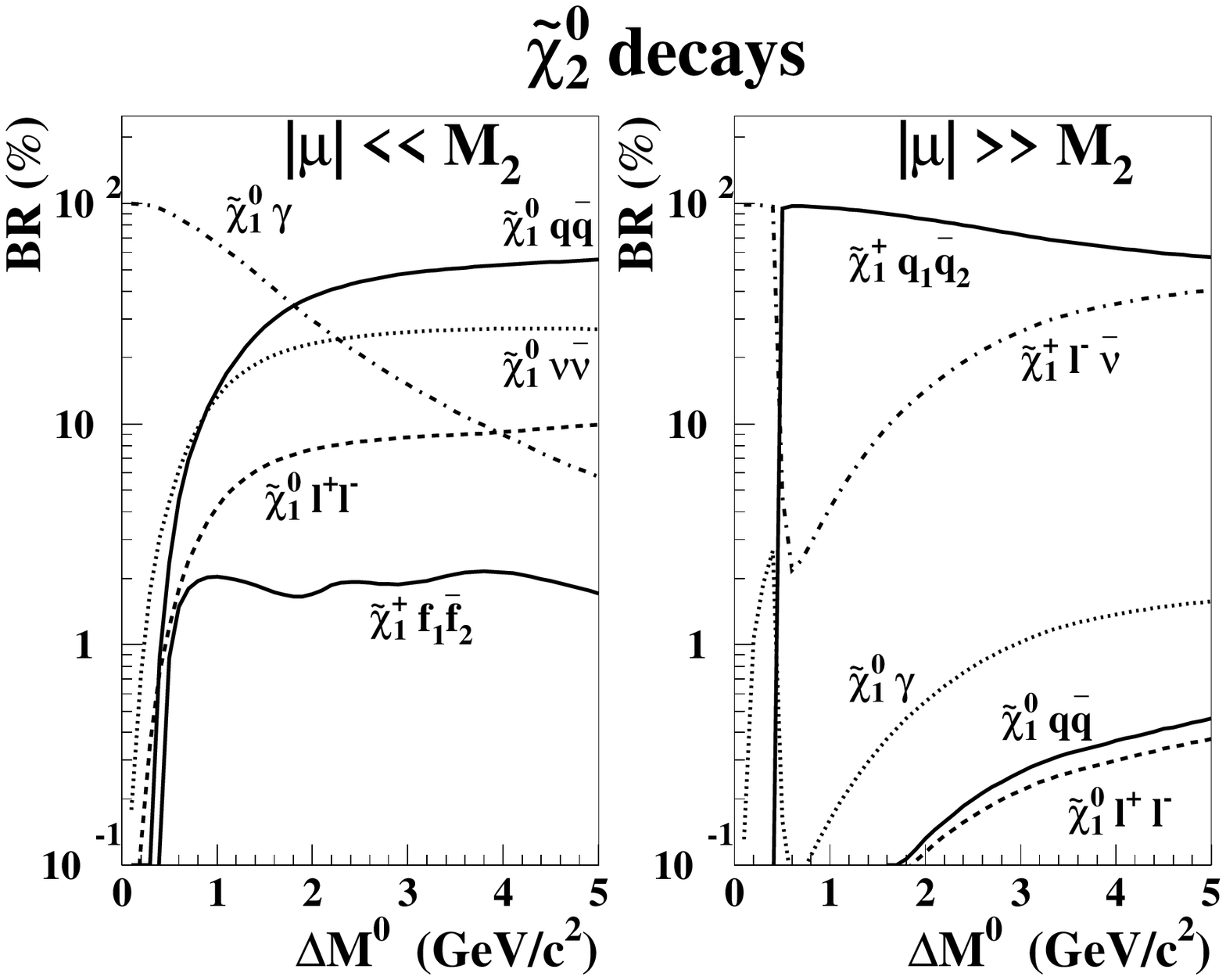} }
\caption[]{ Predicted branching ratios of the decay modes of the second
   neutralino, with $\mcgdz \simeq 50$~\gevc2, as a function of the mass
   difference between the $\cgdz$ and the $\cguz$.
   The left and right plot refer to the higgsino-like and gaugino-like
   scenario respectively.
                     }
\label{fig:brne}
\end{figure}

Figure \ref{fig:brne} shows the exclusive branching ratios of the second
lightest neutralino as a function of $\DMZ$, in the two scenarios which allow
almost degenerate states. Unlike the chargino case, there is now a strong model
dependence of the decay widths. As already mentioned, the only neutralino
scenario with a sufficient cross-section to be searched for is the higgsino
one. In this scenario the rate of the radiative decay $\cgdz\to\cguz\gamma$
increases in the low $\DMZ$ region of interest here.
Thus a large fraction of the $\cgdz$ decays yield either a low energy photon
or a pair of neutrinos, in addition to the LSP. Since it is difficult to
identify photons below about 1~GeV in the detector and distinguish them from
background, both decay modes can be considered invisible for practical
purposes. For these reasons the present work has been limited to charginos,
leaving neutralinos aside.

\section{Search for long-lived charginos}
\label{par:longliv}

Two methods were used to look for charginos with a visible decay length: 
the search for heavy stable charged particles and the search for decay 
vertices inside the detector. Both searches are described in this section.

\subsection{Heavy stable charged particles}

\begin{figure}[htb]
\centerline{
\epsfxsize=14cm\epsffile{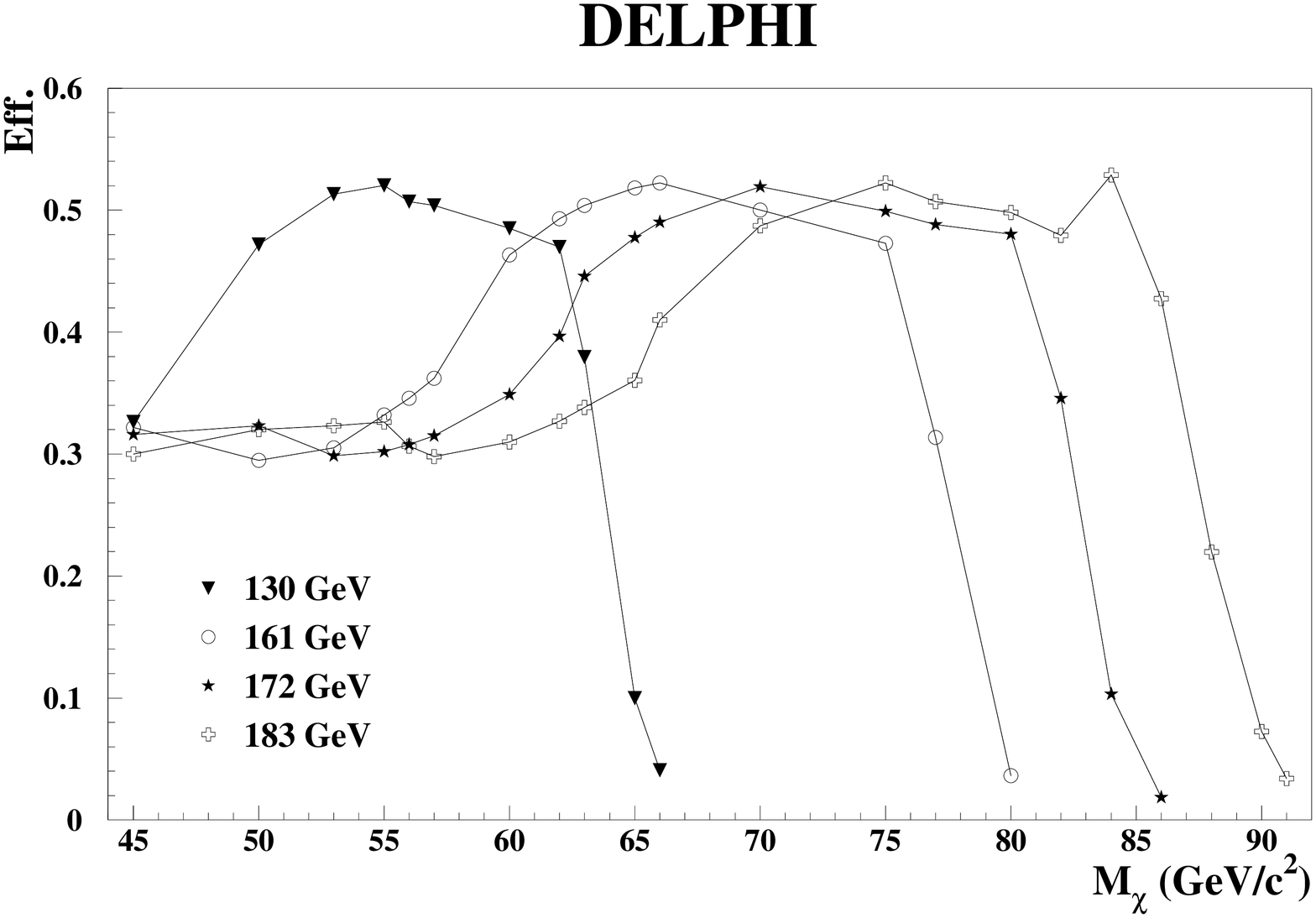}             }
\caption[]{ Efficiency for detecting a single heavy stable chargino in DELPHI
   (produced in pair in the $\epem$ rest frame), as a function of its mass
   and for the different \cms\ energies used in the analysis.
                     }
\label{fig:dedxeff}
\end{figure}

Heavy stable or almost stable charged particles are identified through their
anomalously high specific ionization in the TPC or by the absence of Cherenkov
light produced in the two radiators of the barrel RICH. The leptonic selection
described in \cite{himass} is the one used for the present analysis.
The efficiency of this selection for pair-produced heavy charged particles,
traversing the full depth of the detector, is given in \cite{himass}.
Figure~\ref{fig:dedxeff} shows the efficiency for selecting a single
heavy charged particle as a function of its mass and of the LEP \cms\ energy.
The trigger efficiency for high momentum charged particles crossing the full
depth of the TPC and the RICH is practically unity.

Lower limits on the mass of the chargino have been already published
\cite{himass} under the hypothesis that it decays predominantly outside the
DELPHI detector (in case of an heavy $\snu$ this corresponds to
$\DMP \lsim 100$~\mevc2).
For shorter lifetimes, the detection efficiency can be obtained by convoluting
the efficiency for stable particles with the probability that the chargino
passes through the barrel RICH before decaying.

\subsection{Decay vertices inside the detector (kinks)}

If the heavy charged particle decays inside the central tracking devices of
DELPHI (at a radius between 10~cm and 1~m) then both the incoming and the
outgoing track can be reconstructed, and the angle between the tracks can be
calculated. This method was used in a DELPHI search for scalar tau leptons
decaying into a light gravitino and an ordinary tau lepton \cite{kinks}.
For small $\DMP$, however, the visible momentum of the decay products is quite
small (typically less than $1$ GeV/$c$), and the identification of the
secondary track and, therefore, of the kink becomes more difficult.
The selection criteria adopted in \cite{kinks} have therefore been
modified for the present search, exploiting the typical topology of these
events: two particles emitted in opposite hemispheres decaying into one
low-energy charged particle each.

A set of rather loose general requirements was imposed on the events in order
to suppress the low energy background (beam-gas, beam-wall, etc), two-photon,
$\epem$ and hadronic events:
\begin{itemize}
  \item{there must be at least one charged particle and not more than five;}
  \item{the visible energy must be above 10~GeV;}
  \item{the total energy in electromagnetic showers was required to be
        below 60 GeV;}
  \item{the transverse momentum with respect to the beam axis had to be
        greater than 5~GeV/$c$;}
  \item{the energy measured in the STIC must not exceed 10~GeV.}
\end{itemize}
To compute the above quantities the reconstructed  charged particles were
required to have momenta above 100~MeV/$c$ and impact parameters below 4~cm
in the transverse plane and below 10~cm in the longitudinal direction
(however, no requirement on the impact parameters were imposed to the
reconstructed tracks in the following steps).
Clusters in the calorimeters were interpreted as neutral particles if they
were not associated to charged particles and if their energy exceeded 100~MeV.

All the charged particle tracks were grouped in clusters according to their
measured point closest to the interaction vertex (starting point). The 
clustering procedure is described in \cite{kinks}. Each cluster contains all
tracks whose starting points differ by less than 2~cm. The starting point of
a cluster is defined as the average of the starting points of its tracks. This
procedure allows for clusters with a single track.

A cluster with only one track was considered a $\cgup$ candidate if:
\begin{itemize}
  \item the distance of the starting point from the beam spot, in the plane
    transverse to the beam axis ($xy$ plane),
    $R_{\mathrm{sp}}^{\cgup}$, was smaller than 10~cm;
  \item its momentum was greater than 20~GeV/$c$;
  \item the polar angle of the momentum had to satisfy $|\cos\theta|<0.8$;
  \item the impact parameters of the track along the beam axis and in
    the plane perpendicular to it were less than 10 and 4~cm, respectively.
\end{itemize}

For each single track cluster fulfilling the above conditions, a search was
made for a second cluster with starting point in the transverse plane beyond
$R_{\mathrm{sp}}^{\cgup}$, and an angular separation between the directions
defined by the beam spot and the starting points of the clusters smaller than
$90\deg$ in the $xy$ plane. This secondary cluster was assumed to be formed by
the decay products of the $\cgup$. Therefore, the $\cgup$ candidate and the
secondary cluster had to define a vertex. If the secondary cluster included more
than one track, only the track with the highest momentum was used to search for
the decay vertex or kink (crossing point with the $\cgup$ track).

The crossing point was defined as the midpoint of the line segment connecting 
the points of closest approach in the $xy$ plane between the two (possibly
extrapolated) tracks: the candidate $\cgup$ track and the selected track from
the secondary cluster. The following conditions were required to define a good
crossing point:
\begin{itemize}
  \item the minimum distance between the tracks had to be smaller
        than 1~mm in the $xy$ plane,
  \item the crossing point, the end point of the $\cgup$ track and the
        starting point of its decay products were required to satisfy
        the following conditions:
        \begin{eqnarray}
    -10 \, \mbox{cm} \,\,&<\,\,(R_{\mathrm{cross}}-R_{\mathrm{end}}^{\cgup})\,\,<& \,\,25\, \mbox{cm}
        \nonumber \\
    -25 \, \mbox{cm}\,\,&<\,\,(R_{\mathrm{cross}}-R_{\mathrm{sp}}^{\mathrm{dec.\, prod.}})\,\,<& \,\,10\, \mbox{cm}
        \nonumber
        \end{eqnarray}
        where $R_{\mathrm{end}}^{\cgup}$, $R_{\mathrm{cross}}$
        and $R_{\mathrm{sp}}^{\mathrm{dec.\, prod.}}$
        are the distance from the beam spot to the end point of the $\cgup$
        track, the crossing point of the tracks and the starting point of the
        tracks supposed to come from the decay products of the chargino, in
        the $xy$  plane.
\end{itemize}

Reconstructed secondary vertices could also be the result of particles
interacting in the detector material, or bremsstrahlung, giving a particle
trajectory reconstructed in two separate track segments.
To eliminate this kind of background, events with a good crossing point (kink)
were subjected to additional requirements:
\begin{itemize}
  \item to reject hadronic interactions, any secondary vertex reconstructed
        in the region of the detector where there is material must be outside
        a cone of half opening angle of $5\deg$, with apex at the beam spot
        and centred around the kink direction;
  \item to reject photon radiation, in the case of secondary clusters with
        only one track, no neutral particle was allowed in a $1\deg$ cone
        around the direction of the missing momentum, defined by the difference
        between the momentum of the $\cgup$ and that of the daughter;
  \item to reject segmented tracks, the angle between the tracks
        used to define a vertex, calculated at the crossing point, had to be
        larger than $2\deg$.
\end{itemize}

Finally, for an event to be accepted, at least one charged particle must be 
found in each hemisphere (defined by the plane which contains the beam spot
and is perpendicular to the line connecting the beam spot to the kink).

\begin{figure}[htb]
\centerline{
\epsfxsize=12cm\epsffile{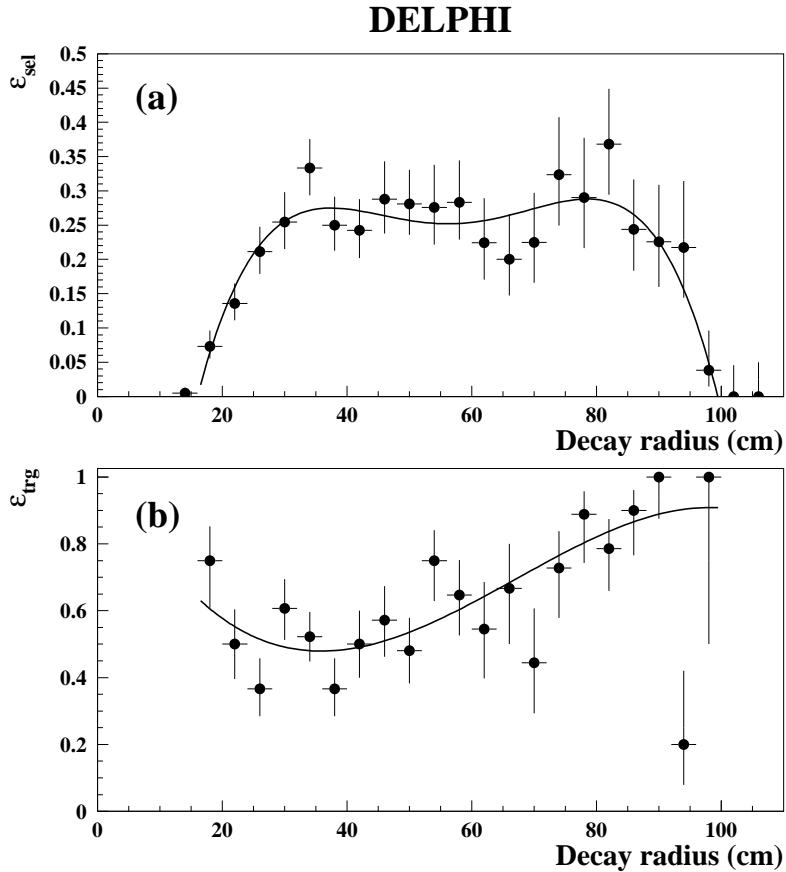}             }
\caption[]{ Efficiency for the kink reconstruction of a $65$ \gevc2 chargino
   with $\DMP \simeq  150$ \mevc2, as a function of  the chargino decay radius,
   at the \cms\ energy of $183$ GeV:
   (a) selection efficiency for the single arm; (b) single particle trigger
   efficiency for the selected kinks. The lines superimposed represent the
   result of the fit described in the text.
                     }
\label{fig:kink}
\end{figure}

This selection was applied to samples of $\epem \to \cgup\cgum$ events,
generated at $\sqrt{s}=161$, $172$ and $183$ GeV and passed through the full
DELPHI simulation and reconstruction chain. The efficiencies for the single
vertex reconstruction as a function of the radial distance from the decay to
the beam spot are plotted in figure~\ref{fig:kink}a, where the case of a 
65~\gevc2 chargino with $\DMP \simeq 150$ \mevc2 has been taken as example.
The selection efficiency ($\esel$) is almost independent of the decay radius
for radii between 30 and 90~cm for almost all the masses generated; it tends
to decrease only when the mass of the chargino approaches the kinematical
limit, because the momenta of the secondaries are not sufficiently enhanced by
the reduced boost. The dependence of the selection efficiency on $\DMP$ is weak
for the values of $\DMP$ within the range searched for with this method. There
is however some increase of the efficiency with increased $\DMP$, as the mean 
momentum of the decay products gets higher.

The trigger efficiency ($\etrg$) depends on the mass of the chargino and on
the \cms\ energy. It has been estimated with a simplified simulation of the
single track trigger in DELPHI, taking the overall trigger efficiency as the
logical OR of the trigger efficiencies of every single charged track in the
event \cite{trigger}. This is a conservative estimate because the possible
contribution of all higher multiplicity triggers was neglected. 
The efficiency for triggering on the hemisphere of the kink, as a function of
the decay radius of the chargino, is shown in figure~\ref{fig:kink}b, for the
same sample as in figure~\ref{fig:kink}a. The radial dependence of both 
selection and trigger efficiencies was fitted with polynomials, allowing a 
different fit at every mass generated. The results of these fits were used in
the analysis and the efficiencies for masses other than those simulated were
obtained by interpolation.

\subsection{Global efficiency for charginos with visible decay length}

The selections in the search for long-lived charginos yield an efficiency
$\eff (x)$ which is a function of the decay radius $x$ of the chargino. Each
chargino produced in the $\epem$ collision can be selected by either one of the
two searches: if it decays outside the RICH, by the search for stable charged 
particles; if it decays inside the ID or the TPC, by the search for kinks. 
The global efficiency, which is a function of the decay radius, is given by the 
logical OR of the two selections.

To extrapolate the efficiency of the event selection to the points in the 
space of the SUSY parameters not fully simulated, a semi-analytic calculation
was used. The distribution of the decay length of the chargino in the reference
frame of DELPHI was derived analytically in the scenarios studied and for any
given chargino mass, $\DMP$ and \cms\ energy. These distributions were 
convoluted with the efficiencies of the experimental search methods $\eff (x)$,
as determined for the fully simulated events, giving the detection efficiency
in any point of the space of the SUSY parameters. 

\begin{figure}[htb]
\centerline{
\epsfxsize=12cm\epsffile{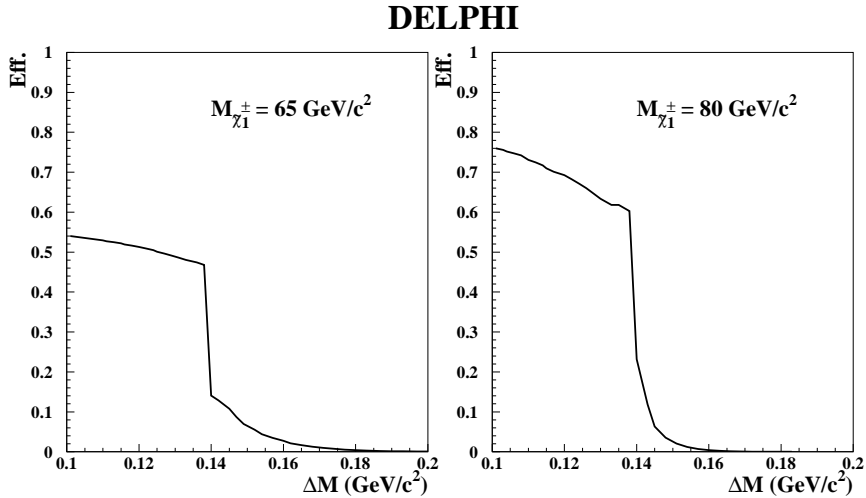}  }
\caption[]{
    Total efficiency for detecting a long-lived charged gaugino in DELPHI using
    the analysis at 183 GeV as function of $\DMP$ (that is of its lifetime) and
    in the approximation of heavy sneutrinos. The two masses of 65 and 80~\gevc2
    have been chosen as examples.
                     }
\label{fig:longliveff}
\end{figure}

As an example, figure~\ref{fig:longliveff} shows the combined detection and
trigger efficiency at 183~GeV for $\epem \to \cgup\cgum$, when the charginos
are pure gauginos and $\msnu=1$~\tevc2, $\mchi=65$ or 80~\gevc2, as a function
of $\DMP$. The efficiencies displayed include both searches for long-lived
charginos (for heavy stable particles and for kinks).

\subsection{Results}

The search for heavy stable particles was performed with all the data collected
at $\sqrt{s}=130$/$136$, $161$, $172$ and $183$ GeV. No candidates remained in
the data, while $0.7 \pm 0.3$ background events were expected \cite{himass}.

In the search for kinks, only data taken in 1996 and 1997 at the \cms\ energies
of $161$, $172$ and $183$ GeV were analysed. No events were selected, while the
background expected at the three \cms\ energies was, respectively,
$0.11 \pm 0.11$~, $0.04^{+0.11}_{-0.04}$ and $0.21\pm 0.14$ events.

In the absence of candidates selected in any of the searches, the $95\%$
confidence level (CL) upper limit is 3.0 events \cite{pdgold} for the whole
statistics analysed. When this limit is considered together with the integrated
luminosities, the expected cross-sections and efficiencies for different SUSY
parameters, the $95\%$ CL exclusion regions in the plane $(\mcgup,\DMP)$ shown
in figure \ref{fig:limit} can be derived.
As far as the SUSY scenarios are concerned, three cases were considered (see
also section \ref{par:scen}; here the results for gauginos are further
subdivided according to the mass of the sneutrino):
\begin{enumerate}
 \item{low $|\mu|$, large $M_{1,2}$ and any mass for the sfermions,
                                                     (higgsino-like);}
 \item{high $|\mu|$, low $M_{2}$ and $\msnu\ge 500$ \gevc2 (gaugino-like);}
 \item{high $|\mu|$, low $M_{2}$ and $\mcgup<\msnu<500$ \gevc2 (gaugino-like).}
\end{enumerate}

In all cases, $\tan{\beta}$ was varied between 1 and 50, $M_2$ and $|\mu|$
between 0 and 100~\tevc2 and $R_f$ between 1 and 10 (although for $R_f<2$ it is 
not possible to have charged gauginos with masses close to that of the lightest
neutralino). The lowest value of the cross-section obtained in the scan of the
SUSY parameters for each scenario, at a given mass of the chargino, was used
for the calculation, so that the limit obtained remains valid for all values
of the parameters. A coarse scan extending to larger values of $R_f$ suggests
that all the results remain valid also for $R_f>10$. The limits obtained in the
third case are much less stringent than the other two, because a light
sneutrino increases the chargino decay width through virtual $\snu$ exchange.
Whenever $\msnu < \mcgup$ the exchanged sneutrino becomes real and the lifetime
drops: this region cannot be excluded by the search for long-lived charged
particles.

Possible systematic biases of these limits have been considered. The relative
statistical errors on the selection and trigger efficiencies propagate at the
second order to the numerical value of the final limit \cite{cousins}; this
is expected to raise by no more than a fraction of a percent the upper limit of
the cross-sections attainable and, for that reason, those errors were not taken
into account further. The efficiencies found in the search for kinks can be
slightly overestimated for $\DMP$ smaller than those used in the full
simulation, because in that case the mean momentum, and so the detectability,
is lower. However, this would not affect the overall limits obtained, as this
region is fully covered by the search for stable particles. Instead, at larger
$\DMP$ the efficiencies can be slightly underestimated; in that case the limits
displayed in figure \ref{fig:limit} are conservative (that is the confidence
level is $95\%$ or higher). The same holds for the trigger efficiency.
In conclusion, the limits obtained are reliable, or conservative, in the whole
space of the SUSY parameters spanned by the present search.

\section{Search for charginos with ISR photons}
\label{par:ISRTAG}

To look for short-lived charginos close in mass to the lightest
neutralino, events with a few low energy particles accompanied by an ISR photon
at high transverse momentum were searched for.
Samples of $\epem \to \cgup\cgum$ events with initial (and final) state
radiation were simulated with $\DMP$= 0.3, 0.5, 1 and 3~\gevc2. The criteria
used to search for such events in the data were then defined on the basis of
these samples and the simulated background samples.

\subsection{Charged and neutral particle selection}

Tracks reconstructed in the detector were considered as charged particles, to
be used in the subsequent analysis, if their momenta were above $100$ MeV/$c$
and known with an error below $100\%$, their impact parameters were below 
$4$~cm in the transverse plane and below $10$~cm in the longitudinal direction
and if the tracks were at least $30$~cm long.

To be accepted, neutral particles had to have an energy of at least $500$ MeV.
Both calorimetric showers and clear photon conversions in the material in front
of the HPC were considered. Photon conversions were identified using the 
standard DELPHI algorithm \cite{delphidet} which looks for a pair of tracks
originating from a common secondary vertex and with an invariant mass 
compatible with zero. 

Subdetector dependent criteria were used to reduce the rate of spurious neutral
particles reconstructed from electronic noise in the calorimeters or, in the
case of the HPC, from $\alpha$-emission by radionuclides embedded in the lead
of the converter material.

In the showers reconstructed by the HPC at least three of the nine layers must
have given a signal; the first layer with a signal must be before the sixth HPC
layer; not more than $90\%$ of the total energy of the shower must be deposited 
in a single layer; the polar angle of the shower axis must point towards the
main event vertex within $15\deg$. 

In the FEMC a signal was required in at least two towers. 
In the data collected since 1997 a more refined quality requirement was used,
defining a frame of $3\times 3$ glasses, centered on the barycentre of the 
shower: a shower was discarded if its energy was above 8~GeV and more than 
$94\%$ of this energy was deposited in the central glass, or if it had no more
than three glasses hit and these were all lined up in the same row or column.

In the STIC an energy of more than 2~GeV must be associated to the shower and
at least two towers must be hit.

All neutral showers inside a cone with half opening angle of $5\deg$ were
combined. The resulting shower was not considered an ISR photon candidate if
the fraction of energy detected in the hadron calorimeter was above $10\%$ of
the total shower energy.

Given the reduced reconstruction and association efficiency of tracks in the
two endcaps as compared to the barrel, spectator electrons from two-photon
events can be seen as showers in the calorimeters without the corresponding
track elements, thus faking neutral particles. Such electrons constitute a
serious background to ISR photons, 
and care must be taken to reject these events.

\subsection{Event selection}

The analyses of the data collected before year 1997 and in 1997 have been done
separately, with slightly different selection criteria. 
The tighter preselection cuts of 1997 were intended to reduce further the
overall volume of the data used in the analysis, given the increased luminosity
collected with respect to the previous years. The final selection of 1997 data
also exploited the improved charged particle rejection in the forward regions
which is obtained by trying to associate hits in the Silicon Tracker to the 
neutral showers.

\subsubsection{Preselection}

In the preselection, events were required to have at least two and at most ten
charged particles.
There should be an isolated photon candidate of at least $4$ GeV, having a
transverse energy above $2$ GeV (above $4.5$ GeV in the 1997 data at the \cms\
energy of 183 GeV) and a mass recoiling against it ($\mopp$, defined by
$\mopp^2 = \ecms^2 - 2 \ecms E_{\gamma}$) of at least $90$ \gevc2.
This photon had to be isolated from any other charged or neutral particle in
the event by $15\deg$ or more. The visible energy of all particles excluding
the photon must not exceed $8\%$ of the available \cms\ energy.
If all the particles in the event were in the same hemisphere inside a cone
centered on the beam axis and with half opening angle of $60\deg$, the event was
discarded; this reduces the background coming from beam-gas or beam-wall
interactions. To 
reject most of the two-photon background, also the fraction of 
the total visible energy within $30\deg$ of the beam axis (excluding the ISR 
photon candidate when inside that cone) was required to be less than $60\%$.

\subsubsection{Selection}
\label{par:selisr}

Following the preselection, in order to optimize signal efficiency and
background rejection, more stringent selection criteria were imposed as
follows.
\begin{itemize}
\item{There must be at least two and at most six accepted charged particles
 and, in any case, not more than ten tracks in the event.}
\item{To reject the bulk of the two-photon background, the transverse energy
 of the ISR photon was required to be greater than $(E_{\gamma}^T)^{\mmin}$,
 defined in eq.~(\ref{eq:etgmin}). Here $\theta_{\mmin}$ was taken as the
 minimum angle at which the photon shower is fully reconstructed in the STIC
 ($1.82\deg$). }
\item{$\mopp$ must be above $96$ \gevc2. This was intended mainly to reduce the
 number of events with an on-shell $\Zzero$ recoiling against the photon.}
\item{The photon was required to be isolated by at least $30\deg$ with respect
 to any other charged or neutral particle in the event.}
\item{The sum of the energies of the particles emitted within $30\deg$ of the
 beam axis ($E_{30}$) was required to be less than 25\% of the total visible
 energy. The photon was not considered in any of the two energy sums if its
 direction was inside that cone.}
\item{If the ISR photon candidate was detected in the STIC, it must not be
 correlated with a signal in the Veto Counters.}
\item{In the data collected during 1997, if the ISR photon candidate was at an
 angle between $10\deg$ and $25\deg$ from the beam, the region where the TPC
 cannot be used in the tracking, it must not be correlated with hits in the
 Silicon Tracker.}
\item{The visible energy of the event, excluding the photon, must be below
 $5\%$ of $\ecms$. For $\DMP<1$~\gevc2 this fraction was reduced to $2\%$.}
\end{itemize}

\begin{figure}[p]
\centerline{
\epsfxsize=13.5cm\epsffile{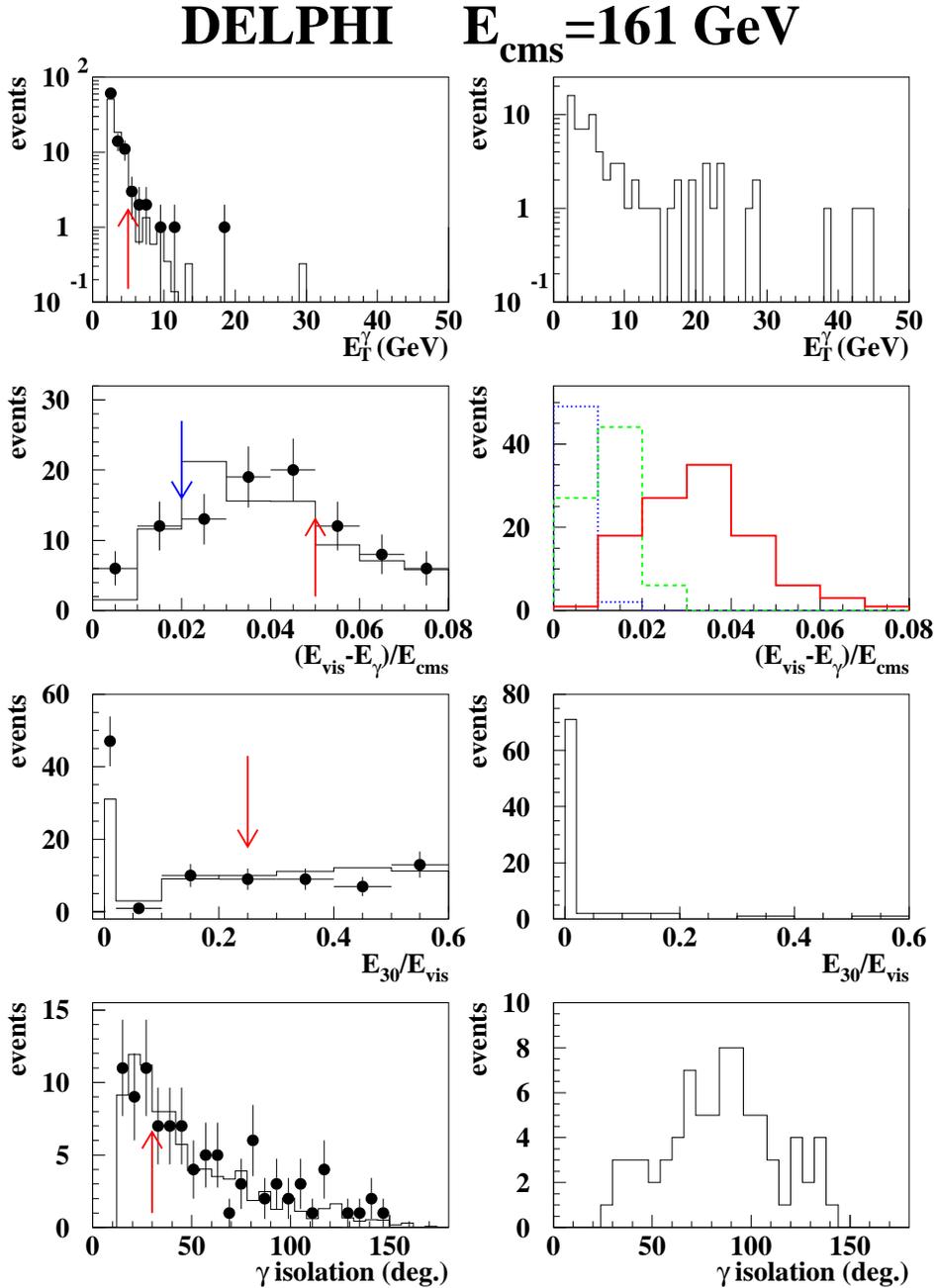} }
\caption[]{ Some of the variables used in the selection at $161$ GeV.
    In the left plots the data (dots) are compared with the SM expectations.
    On the right, as an example, the corresponding distributions (with
    arbitrary normalization) are shown for the signal with $\mcgup = 50$~\gevc2
    and $\DMP =1$~\gevc2. In the plot of the visible energy (second row) 
    all three mass splittings are shown: dotted, $\DMP =0.3$~\gevc2; dashed,
    $\DMP =1$~\gevc2; full line, $\DMP =3$~\gevc2.
                     }
\label{fig:bck161}
\end{figure}

\begin{figure}[p]
\centerline{
\epsfxsize=13.5cm\epsffile{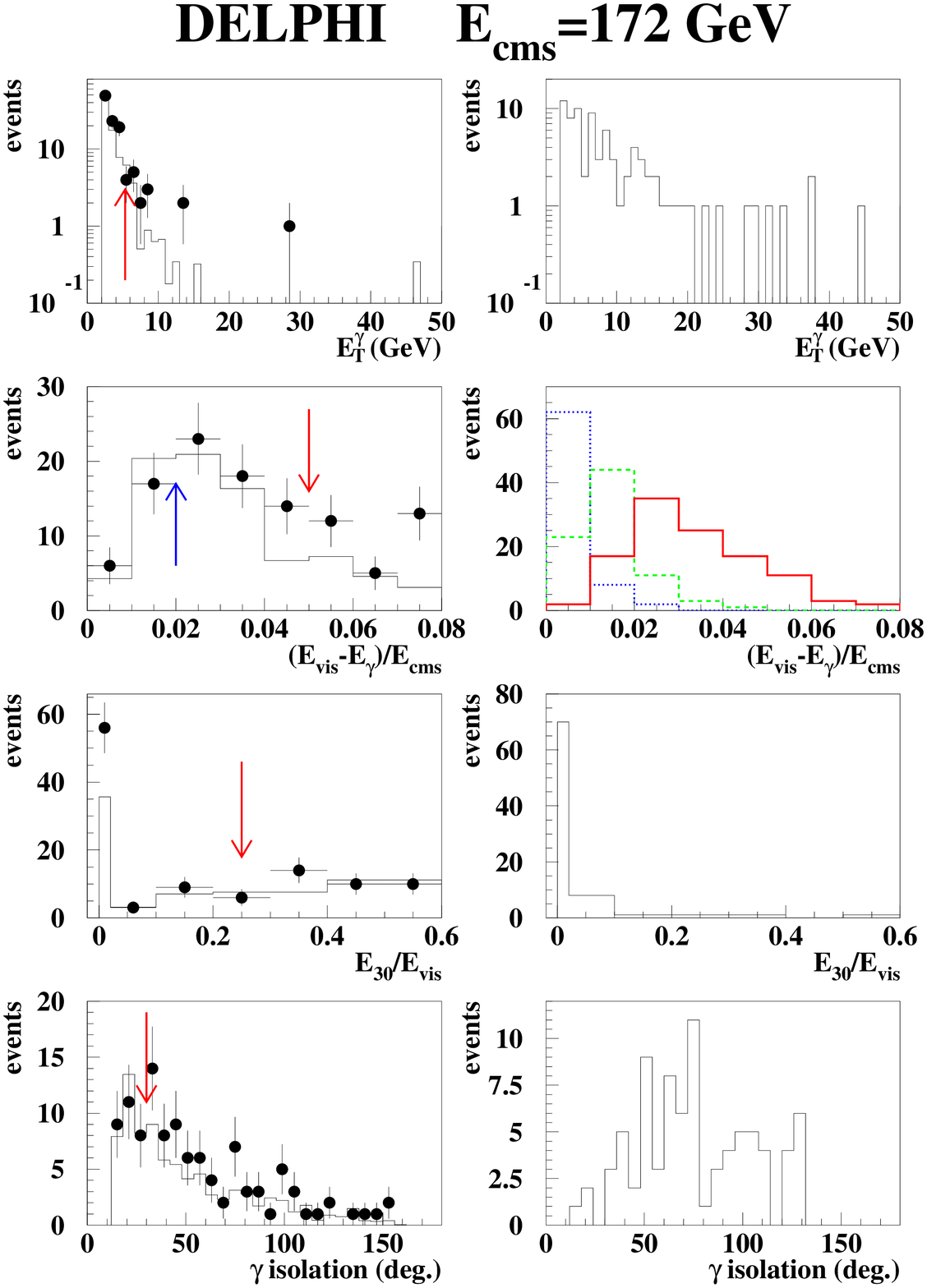} }
\caption[]{ Same as figure \ref{fig:bck161}, but data and simulation refer
    to the samples at $172$ GeV.
                     }
\label{fig:bck172}
\end{figure}

\begin{figure}[p]
\centerline{
\epsfxsize=13.5cm\epsffile{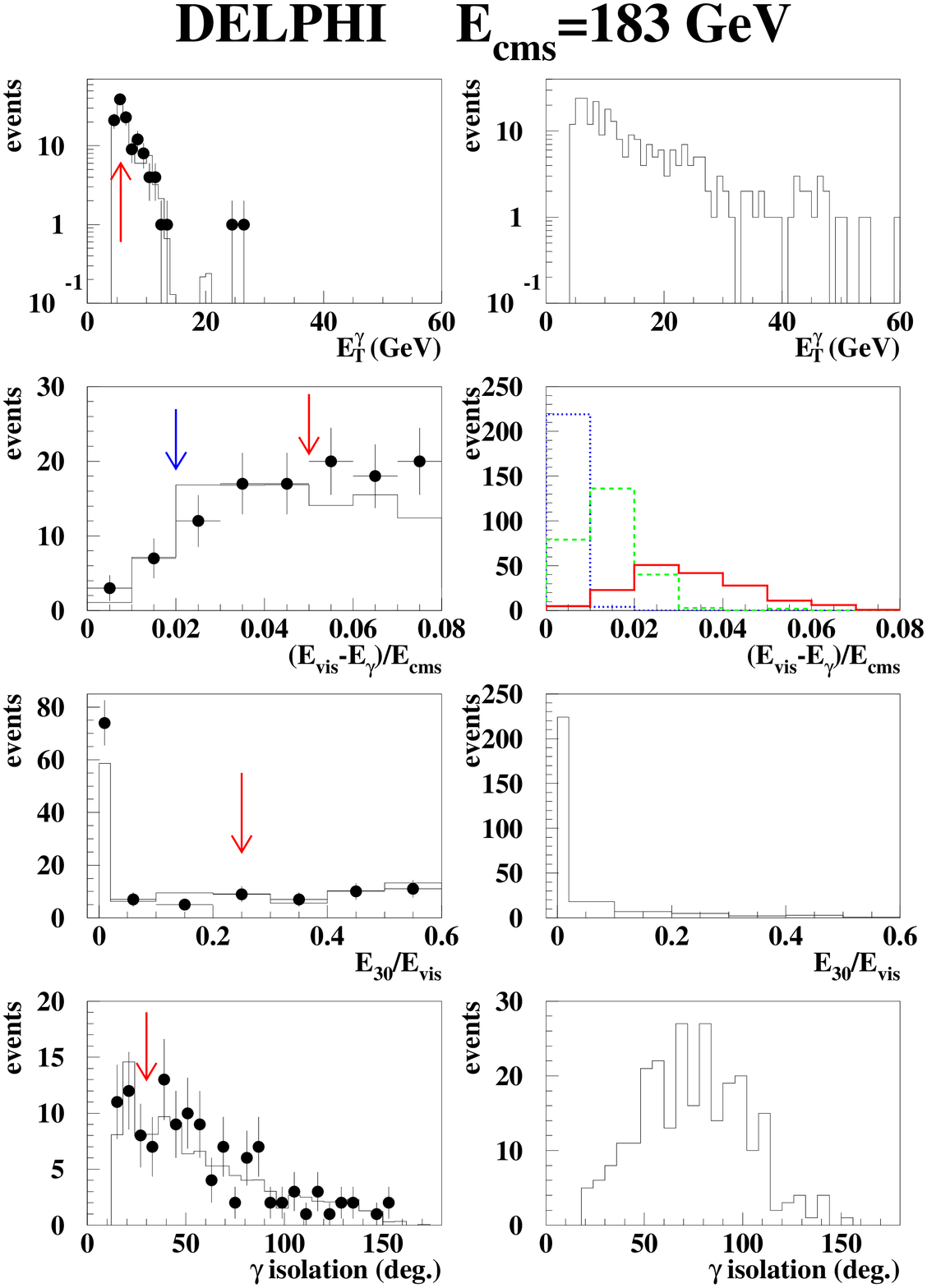} }
\caption[]{ Same as figure \ref{fig:bck161}, but data and simulation refer
    to the samples at $183$ GeV. The cut on the transverse energy of the photon
    in the preselection is tighter than the one used in the previous plots at
    lower \cms\ energies.
                     }
\label{fig:bck183}
\end{figure}

Figures \ref{fig:bck161}, \ref{fig:bck172} and \ref{fig:bck183} show the
distributions of some of the variables used for the selection in the preselected
samples at 161, 172 and 183 GeV respectively. The data are compared with the
SM expectation, normalized to the same luminosity. The agreement is good in 
almost all distributions. The simulated two-photon interactions giving hadrons
via the QPM process with and without transverse momentum ($p_T$) of the ISR
photon were compared. It was evident that the small excess of real data
in the first bins of the distributions of $E_{30}$ and $(\evis-E_{\gamma})$ 
can be, at least qualitatively, explained by the lack of high $p_T$ ISR in 
some of the simulated two-photon samples. At present there are no generators
available, however, which correctly describe the ISR transverse momentum for
these processes.

The corresponding distributions for the signal with $\mcgup = 50$ \gevc2 and
$\DMP = 1$ \gevc2, taken as an example, are shown to the right in the same
figures. The histograms of the visible energy are shown for the three mass
differences of $0.3$, $1$ and $3$ \gevc2, since the energy of the visible
decay products depends on $\DMP$.

For the background estimates at $\sqrt{s}=130$ and $136$ GeV, the problems
with the simulated samples were more important: many of them were originally
generated with insufficient statistics or with stricter requirements than those
used in the subsequent analysis. No detailed comparison of the distributions of
the data with respect to the simulation was meaningful at those energies.

\subsection{Results}

The results of the selection, when applied to data and simulated background
at all the \cms\ energies, are shown in table \ref{tab:cuts}. Six candidate
events remain in the data after the selection for $1\le\DMP\le 3$~\gevc2
(one in the sample at 136~GeV collected in 1995, one at 172~GeV and four at 
183~GeV). Some of their properties are summarized in table \ref{tab:cand}. 
Only two of them pass also the stricter requirements for smaller $\DMP$ (both
at 183 GeV). There is no significant excess above the SM expectations in 
either selection. The SM background remaining at the end of the selection
is almost entirely composed of two-photon interactions.

\begin{table}[thbp]
{ \small
\begin{center}
\begin{tabular}{|c|c|c|c|}
 \hline
 \multicolumn{2}{|c|} {        } & \multicolumn{2}{|c|} {        } \\
 \multicolumn{2}{|c|}
     {\boldmath $ \evis-E_{\gamma} <5\% \cdot \ecms  $ } &
 \multicolumn{2}{|c|}
     {\boldmath $ \evis-E_{\gamma} <2\% \cdot \ecms  $ } \\
 \multicolumn{2}{|c|}
     {\bf \boldmath ($1 \le \DMP \le 3$ \gevc2)   } &
 \multicolumn{2}{|c|}
     {\bf \boldmath ($\DMP < 1$ \gevc2) } \\
 \multicolumn{2}{|c|} {        } & \multicolumn{2}{|c|} {        } \\
            &                    &             &                        \\
      \makebox[6em] {\bf  Data                 }  &
      \makebox[9em] {\boldmath \bf $\Sigma$ backgrounds  }  &
      \makebox[6em] {\bf  Data                 }  &
      \makebox[9em] {\boldmath \bf $\Sigma$ backgrounds  }  \\
            &                    &             &                        \\
 \hline
 \hline
 \multicolumn{4}{|c|} {                     } \\
 \multicolumn{4}{|c|} {
 {\boldmath \bf $\ecms=130$/$136$ GeV} \,\,\,\,\,\,(~$\int {\cal L} = 11.7$ pb$^{-1}$) } \\
 \multicolumn{4}{|c|} {                     } \\
 \hline
            &                    &            &              \\
   $1$      & $0.84 \pm 0.84 $   &       $0$  & $\simeq 0$   \\
            &                    &            &              \\
 \hline
 \hline
 \multicolumn{4}{|c|} {                     } \\
 \multicolumn{4}{|c|} {
 {\boldmath \bf $\ecms=161$ GeV} \,\,\,\,\,\,(~$\int {\cal L} = 9.7$ pb$^{-1}$) } \\
 \multicolumn{4}{|c|} {                     } \\
 \hline
            &                    &            &                \\
   $0$      & $1.12\pm 0.38$     &   $0$      & $0.45\pm 0.21$ \\
            &                    &            &                \\
 \hline
 \hline
 \multicolumn{4}{|c|} {                     } \\
 \multicolumn{4}{|c|} {
 {\boldmath \bf $\ecms=172$ GeV} \,\,\,\,\,\,(~$\int {\cal L} = 9.9$ pb$^{-1}$) } \\
 \multicolumn{4}{|c|} {                     } \\
 \hline
            &                    &            &                \\
   $1$      & $0.64\pm 0.18$     &   $0$      & $0.11\pm 0.06$ \\
            &                    &            &                \\
 \hline
 \hline
 \multicolumn{4}{|c|} {                     } \\
 \multicolumn{4}{|c|} {
 {\boldmath \bf $\ecms=183$ GeV} \,\,\,\,\,\,(~$\int {\cal L} = 50.0$ pb$^{-1}$) } \\
 \multicolumn{4}{|c|} {                     } \\
 \hline
            &                    &            &                \\
   $4$      & $2.96\pm 0.88$     &   $2$      & $0.44\pm 0.21$ \\
            &                    &            &                \\
 \hline
\end{tabular}
\end{center}
 \caption[]{ Results of the selection on the data and on the sum of the
 expected SM backgrounds. The integrated luminosity is the one used for the
 analysis. }
\label{tab:cuts}
 }
\end{table}

\begin{table}[htbp]
{ \small
\begin{center}
\begin{tabular}{|c|c|c|c|c|c|c|c|}
 \hline
         &            &            &            &           &            &            &           \\
    \makebox[3.6em]{ $\ecms$ }                   &
    \makebox[3.6em]{ $N_{\mathrm{charged}}$ }    &
    \makebox[3.6em]{ $N_{\mathrm{neutral}}$ }    &
    \makebox[3.6em]{ $E_{\mathrm{vis}}$ }        &
    \makebox[3.6em]{ $E_{\gamma}$ }              &
    \makebox[3.6em]{ $\theta_{\gamma}$ }         &
    \makebox[4.8em]{ ${\mathrm{IP}}^{r\phi}_1$ } &
    \makebox[4.8em]{ ${\mathrm{IP}}^{r\phi}_2$ }    \\
 (GeV)   &            &            &  (GeV)     &  (GeV)     &  (deg.)    &  (cm)     &  (cm)     \\
         &            &            &            &           &            &            &           \\
 \hline
         &            &            &            &           &            &            &           \\
  136    &    2       &     2      &  40.1      &  34.1     &  165.5     &  $-0.18 \pm 0.03$  &  $~~0.62 \pm 0.06$   \\
  172    &    2       &     2      &  12.6      &   7.9     &   56.8     &  $~~0.00 \pm 0.01$  &  $~~0.00 \pm 0.01$   \\
  183    &    2       &     1      &  54.3      &  53.4     &   10.6     &  $-0.05 \pm 0.03$  &  $~~0.01 \pm 0.02$   \\
  183    &    2       &     1      &  11.7      &   8.4     &   52.2     &  $-0.01 \pm 0.03$  &  $-0.01 \pm 0.01$   \\
  183    &    2       &     2      &  13.2      &   7.0     &  124.9     &  $~~0.05 \pm 0.04$  &  $~~0.00 \pm 0.01$   \\
  183    &    2       &     1      &  13.2      &   5.8     &   85.5     &  $~~0.00 \pm 0.01$  &  $~~0.00 \pm 0.01$   \\
         &            &            &            &           &            &            &           \\
 \hline
\end{tabular}
\end{center}
 \caption[]{ Some of the properties of 
             the events remaining in the data after the selection: 
             the \cms\ energy, the number of charged
             and neutral particles in the event, the energy and the polar angle
             of the photon, the impact parameters in the plane $r\phi$ of the
             two charged tracks.}
\label{tab:cand}
 }
\end{table}

At 130/136 GeV some of the background samples were either missing or had 
requirements at the generation level stricter than those used later in the
analysis and the background quoted in the first line of table~\ref{tab:cuts}
is likely to be underestimated.

\begin{figure}[tbh]
\centerline{
\epsfxsize=12cm\epsffile{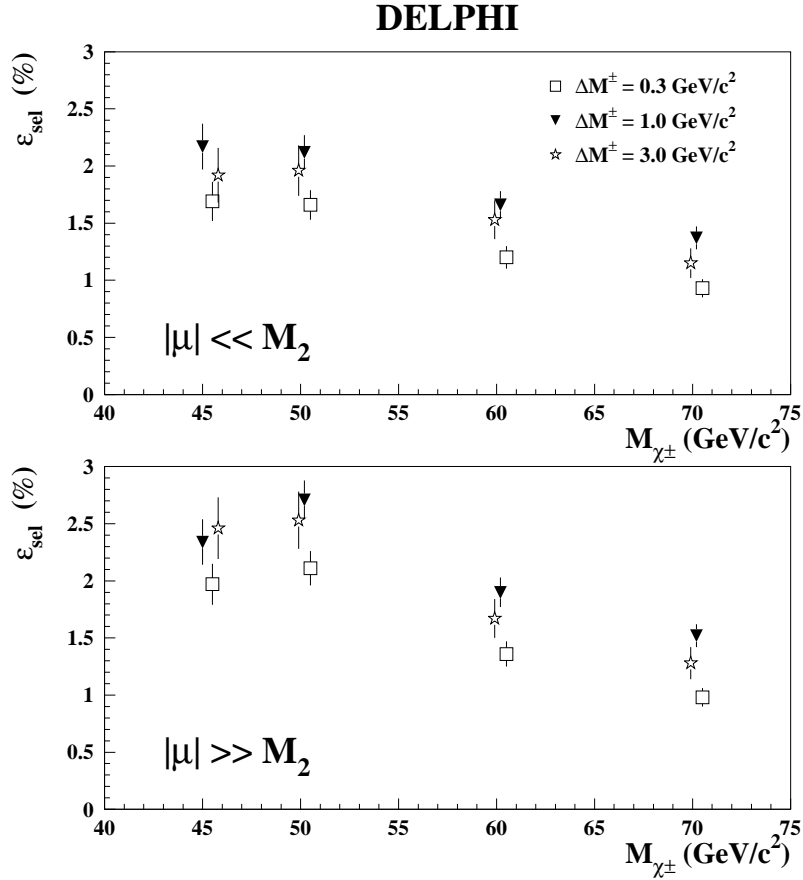} }
\caption[]{ Selection efficiencies for charged higgsinos and gauginos at
 183~GeV as a function of their mass and mass difference with the lightest
 neutralino. This gaugino efficiency is valid only in the approximation of
 heavy sneutrinos; for light sneutrinos a different value is used.
                     }
\label{fig:isreff}
\end{figure}

The selection efficiency ($\esel$) for charged higgsinos and gauginos at 
$\sqrt{s}=183$~GeV, determined by using the samples of simulated events is 
shown in figure \ref{fig:isreff} as a function of the mass of the chargino and
of $\DMP$. Similar efficiencies have been obtained at the other \cms\ energies.
It must be stressed that the very low signal efficiency comes from the 
requirement of having an energetic ISR photon radiated at visible angles. 
As the mass of the chargino increases, the energy of the photon decreases, thus
lowering the overall event selection  efficiency. The selection efficiency for
gauginos, in the case of a heavy sneutrino, is higher than for higgsinos, even
though the decays are
similar in the two scenarios, because the gaugino cross-section resonates more
strongly around the $\Zzero$ pole, leading to more frequent ISR radiation.
However, the gaugino efficiency is significantly smaller than the one shown in
the figure when the mass of the sneutrino is close to the mass of the chargino:
light sneutrinos would enhance the fraction of leptonic decays, which have
additional missing energy and then a lower efficiency for the same $\DMP$.
All this was taken into account when computing the exclusion limits.

Below transverse momenta of $1$-$2$ GeV/$c$ the single track trigger efficiency
of DELPHI starts to decrease \cite{trigger}. The events searched for have few
tracks of low energy (besides the energetic photon); for that reason, a
dedicated and detailed study of the trigger performances at low visible
transverse energy was carried out. The trigger efficiencies for events with a
visible transverse energy similar to the one of the system accompanying the
ISR photon in the signal samples which pass the selection cuts, were computed
from all the data collected at LEP2. This study had to be repeated for every new
year of data taking, since the definitions of the decision functions and the 
downscaling factors applied in case of noise during the data acquisition can
affect the results. The sample of events with independent trigger signals from
any of the calorimeters was used to compute the efficiency for the tracking
components of the trigger. The overall trigger efficiency for the signal 
($\etrg$) was then estimated as the logical OR of the single photon trigger
efficiency \cite{singleg} and the trigger efficiency for events in which there
are only low energy tracks. For the
signal events considered here, the estimated overall trigger efficiency varies
between $82\%$ and $98\%$.

\subsection{Exclusion limits in the search with the ISR tag}

Since there was no evidence for a signal at any of the \cms\ energies studied,
limits were derived on the pair production of charginos nearly mass-degenerate
with the LSP.

\subsubsection{Method}

Although the cross-section is relatively high for all chargino masses (see
figure \ref{fig:xps}), it is important to combine the data taken at all
energies, because of the low signal efficiency. The method used to obtain a
combined limit from the data taken at the different \cms\ energies
is the following.

The number of candidates observed at each \cms\ energy is distributed according
to a Poissonian. The (Bayesian) probability density of the true mean value for
the number of signal events in the data ($\nsig$) is derived according to
\cite{obr}, given the number of observed events and the expected number of
background events at each $\ecms$. The error on the expected background
content was taken into account by assigning a Bayesian probability density to
different values of the background and weighting the $\nsig$ densities by the
background density. The background densities used were taken to be Gaussian at
every $\ecms$, with standard deviations equal to the errors reported in
table~\ref{tab:cuts}.

The total number of expected signal events is
\begin{equation}
\label{eq:nexp}
  \nexp =  \sum \sigma_i {\cal L}_i \, \esel \etrg
\end{equation}
where $i$ runs over all the \cms\ energies. The statistical errors on the
selection efficiencies completely dominate over the errors on the integrated
luminosities and trigger efficiencies, which were neglected. For the
cross-sections, the lowest values obtained in the scan of the SUSY parameters
were used. $\nexp$ was then assigned a Bayesian probability density, based on
the binomial statistics relevant in the calculation of $\esel$, and assuming
equal a priori probabilities.

The two probability densities for $\nsig$ and $\nexp$ are independently
determined, and the probability that $\nsig < \nexp$ can therefore be obtained
by convoluting these two densities using Monte-Carlo techniques. If this
probability, in a given scenario, is equal to $\eta$ for a point in the plane
$(\mcgup,\DMP)$, then the point is excluded at the confidence level $\eta$.

In this way, confidence levels of exclusion were derived for the mass points
where a full simulation of the signal had been performed. Between these points
an interpolation, based on SUSYGEN events without full detector simulation,
was used to obtain the limit.

\subsubsection{Limits}

Table \ref{tab:lim} gives the $95\%$ CL lower limits on the mass of the
chargino for each of the three scenarios considered and for the different
$\DMP$ ranges. To compute the efficiencies, event samples have been generated
at fixed values of the mass of the chargino and of $\DMP$. 
While it is straightforward to interpolate between simulated points at the
same $\DMP$ and different $\mcgup$, it is more difficult in the case of
different $DMP$ because of the different $Q$-values in the chargino decay.
For this reason, a limit was calculated for each $\DMP$ simulated by 
interpolation in $\mcgup$, but no interpolation in $\DMP$ was done.
Instead, the limit between any two simulated $\DMP$ was conservatively
taken as the the lower of the corresponding two $\mcgup$ limits, giving a
step-like exclusion contour. In particular, no limit was derived below
the minimum $\DMP$ used in the simulation; however, using a small sample of
simulated events it has been verified that those efficiencies drop quickly
below that lowest $\DMP$, at the \cms\ energies studied in this paper.

\begin{table}[tbhp]
{ \small
\begin{center}
\begin{tabular}{|c|c|c|c|}
 \hline
         &                        &                    &                   \\
         &  \makebox[7.0em]{$|\mu|\ll M_2$}
         &  \makebox[7.0em]{$|\mu|\gg M_2$}
         &  \makebox[7.0em]{$|\mu|\gg M_2$}               \\
         &                        & $\msnu > 500$ \gevc2
                                  & Any $\msnu > \mcgup$  \\
         &                        &                    &                   \\
 \hline
            &                        &                    &                \\
 $0.3 \le \DMP    < 0.5$ \gevc2
            &   48.0 \gevc2          &  62.6 \gevc2       &    -           \\
            &                        &                    &                \\
 $0.5 \le \DMP    < 1.0$ \gevc2
            &   48.0 \gevc2          &  62.6 \gevc2       & 49.4 \gevc2    \\
            &                        &                    &                \\
 $1.0 \le \DMP  \le 3.0$ \gevc2
            &   49.9 \gevc2          &  60.6 \gevc2       & 48.2 \gevc2    \\
            &                        &                    &                \\
 \hline
\end{tabular}
\end{center}
            }
 \caption[]{ $95\%$ CL lower limits on the mass of the chargino obtained with
 the search for soft particles accompanied by an energetic ISR photon, in the
 three scenarios in which a mass-degeneracy with the neutralino is possible.}
\label{tab:lim}
\end{table}

When the chargino is a gaugino and the mass of the sneutrino is above but
close to $\mcgup\,$, the decay $\cgup \to \cguz l^+ \nu_l$~, mediated by a
virtual $\snu$, is enhanced with respect to the branching ratios shown in
figure~\ref{fig:brch}. In that case there are two invisible particles which
carry away the energy and the visible lepton is usually softer than the pion
of the two-body decay $\cgup \to \cguz \pi^+$. 
For this reason the selection efficiency becomes lower with a light sneutrino
with mass above the chargino mass and the limit has been set only down to 
$\DMP=500$~\mevc2 in table~\ref{tab:lim}.
If the sneutrino is lighter than the chargino, whether or not it is the LSP,
the $\Delta M$ to be considered is the mass difference between the chargino
and the sneutrino: if $\msnu>\mcguz$ the subsequent decay $\snu \to \nu \cguz$
is experimentally invisible; if $\msnu\le\mcguz$ the chargino nevertheless
decays predominantly into $l\snu$.

All these results take into account a variation of $\tan \beta$ between $1$
and $50$ and a variation of the $M_1$, $M_2$ and $\mu$ parameters so that the
mass difference between the chargino and the neutralino remains below
$3$~\gevc2 and $M_2 \le 2M_1 \le 10M_2$ (although a coarse scan of the space of
the SUSY parameters indicates that all limits would remain the same even if
$M_1>5M_2$).

As anticipated in section \ref{par:data}, the limitations of some of the
two-photon generators used lead to an underestimate of the SM background in the
region of high $p_T$ ISR where the signal is expected. There is no evidence of
any signal, however, and when deriving exclusion limits the only bias expected
because of the somewhat inadequate simulation is that the mass limits are
likely to be underestimated. In other words, the limits are conservative, in
the sense that with a more precise background simulation their confidence level
would be likely to exceed $95\%$. No attempt to compensate for this effect has
been made.

\begin{figure}[tbhp]
\centerline{
\epsfxsize=15.5cm\epsffile{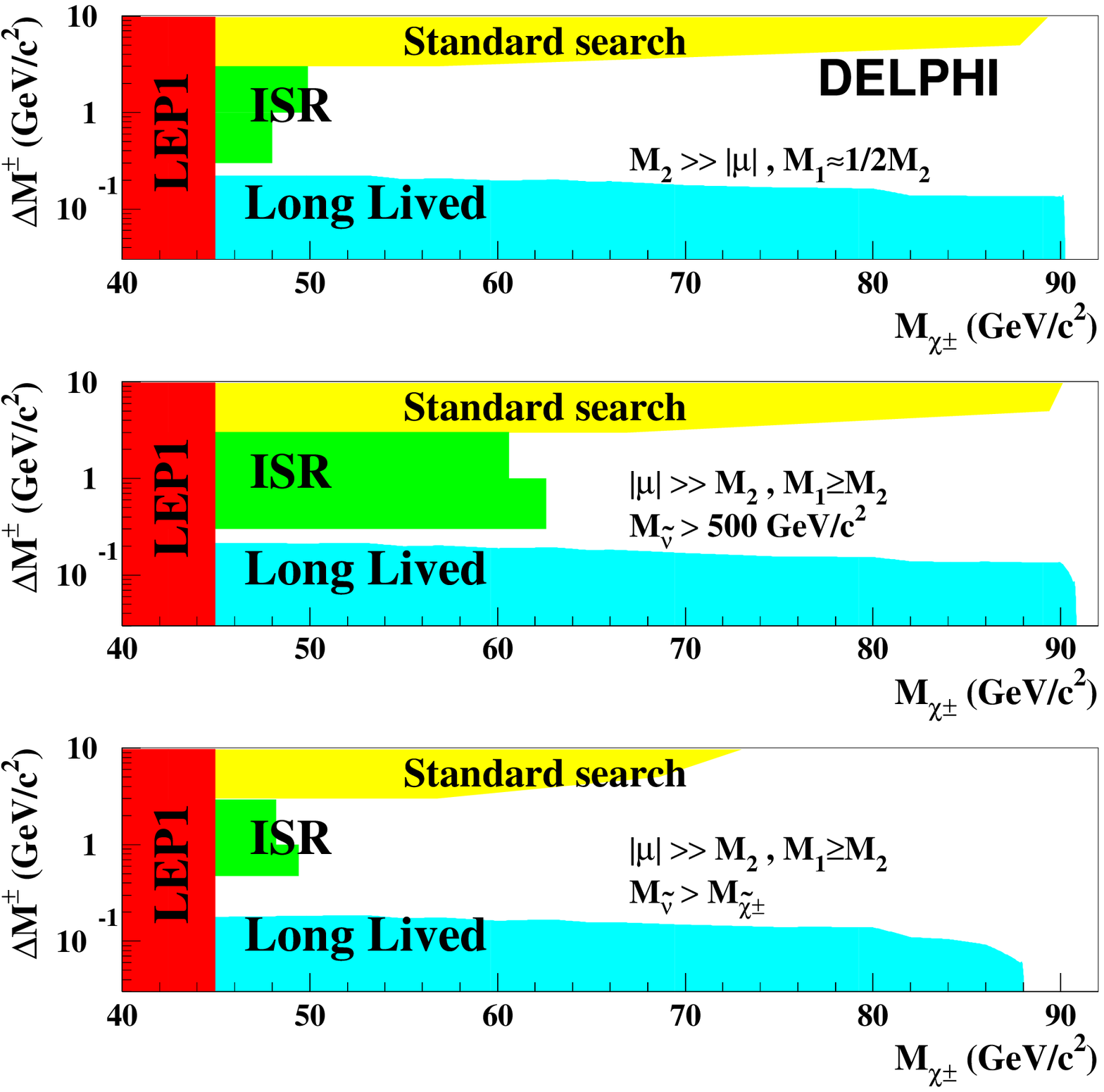} }
\caption[]{ Regions in the plane ($\mcgup$,$\DMP$) excluded by DELPHI at
  the $95\%$ CL using: the search for high $\DMP$ charginos; the search for
  soft particles accompanied by ISR; the search for long-lived charginos.
  The three scenarios are the ones which allow low $\DMP$: the lightest
  chargino is a higgsino; the lightest chargino is a gaugino and
  $\msnu>500$ \gevc2; the lightest chargino is a gaugino and $\msnu\ge\mcgup$
  (or $\msnu<\mcgup$ and $\mcgup-\msnu$ replacing $\DMP$ on the $y$ axis in the
  search with the ISR signature).
                     }
\label{fig:limit}
\end{figure}

Figure \ref{fig:limit} shows these limits together with the ones obtained in
the search for long-lived charginos. For completeness, also shown are the
limits obtained at LEP1 \cite{pdgnew,lep1} and the results of the search
for high $\DMP$ charginos in DELPHI \cite{delphicharg}.

\section{Conclusions}

Charginos nearly mass-degenerate with the lightest neutralino (assumed to be
the LSP) have been searched for using the data collected by the DELPHI
experiment during the runs of LEP above the $\Zzero$ pole energy from 1995 to
1997. Two different approaches were used.

For extremely small mass differences ($\DMP \le 200$ \mevc2) the information
contained in the long lifetime of the chargino (decay length) was exploited.
The search for
heavy particles decaying outside the central subdetectors of DELPHI used the
specific ionization of the tracks in the TPC and the light cone produced in
the RICH detector. Chargino decays, between about $15$~cm and $1$~m, would be
seen as kinks in the tracks reconstructed in the central tracking detectors of
DELPHI. No candidate events were found in the data collected at $130$, $136$,
$161$, $172$ and $183$~GeV using the first analysis, or at $161$, $172$ and
$183$~GeV using the second analysis.

When the lifetime is so short that the decay vertex cannot be seen inside the
sensitive devices of DELPHI and $\DMP$ is too small ($0.3<\DMP<3$ \gevc2) to be
selected by the usual criteria adopted in the search for charginos, some events
can still be recovered by looking for the typical topologies of the chargino
decays at low $\DMP$ accompanied by a high energy photon radiated from the
initial state. The ISR signature reduces the otherwise overwhelming two-photon
background to acceptable rates, although it also strongly affects the signal
efficiency. It is necessary to combine all the statistics collected so far at
the different LEP2 \cms\ energies to achieve sufficient sensitivity.
No evidence of a signal has been found in the data collected by DELPHI at the
\cms\ energies of $130$, $136$, $161$, $172$ and $183$ GeV.

The regions of the plane $\mcgup$ vs. $\DMP$ excluded at the $95\%$~CL by the
combination of these searches in DELPHI are summarized in
figure~\ref{fig:limit}. There is still an inaccesible region at, approximately,
200~\mevc2~$<\DMP<$~300~\mevc2. With the higher statistics available and the
increased boost of the decay products of the chargino, provided by the raised
\cms\ energy, this region is likely to be explored with the use of data from
1998 onwards.

\subsection*{Acknowledgements}
\vskip 3 mm
 We are greatly indebted to our technical 
collaborators, to the members of the CERN-SL Division for the excellent 
performance of the LEP collider, and to the funding agencies for their
support in building and operating the DELPHI detector.\\
We acknowledge in particular the support of \\
Austrian Federal Ministry of Science and Traffics, GZ 616.364/2-III/2a/98, \\
FNRS--FWO, Belgium,  \\
FINEP, CNPq, CAPES, FUJB and FAPERJ, Brazil, \\
Czech Ministry of Industry and Trade, GA CR 202/96/0450 and GA AVCR A1010521,\\
Danish Natural Research Council, \\
Commission of the European Communities (DG XII), \\
Direction des Sciences de la Mati$\grave{\mbox{\rm e}}$re, CEA, France, \\
Bundesministerium f$\ddot{\mbox{\rm u}}$r Bildung, Wissenschaft, Forschung 
und Technologie, Germany,\\
General Secretariat for Research and Technology, Greece, \\
National Science Foundation (NWO) and Foundation for Research on Matter (FOM),
The Netherlands, \\
Norwegian Research Council,  \\
State Committee for Scientific Research, Poland, 2P03B06015, 2P03B03311 and
SPUB/P03/178/98, \\
JNICT--Junta Nacional de Investiga\c{c}\~{a}o Cient\'{\i}fica 
e Tecnol$\acute{\mbox{\rm o}}$gica, Portugal, \\
Vedecka grantova agentura MS SR, Slovakia, Nr. 95/5195/134, \\
Ministry of Science and Technology of the Republic of Slovenia, \\
CICYT, Spain, AEN96--1661 and AEN96-1681,  \\
The Swedish Natural Science Research Council,      \\
Particle Physics and Astronomy Research Council, UK, \\
Department of Energy, USA, DE--FG02--94ER40817. \\

\end{document}